# Architectural Design Space for Modelling and Simulation as a Service: A Review

Mojtaba Shahin [a], M. Ali Babar [b, c], Muhammad Aufeef Chauhan [b, c]

[a] Department of Software Systems and Cybersecurity, Faculty of Information Technology, Monash University, Australia

[b] Centre for Research on Engineering Software Technologies, School of Computer Science, The University of Adelaide, Australia

[c] Cyber Security Cooperative Research Centre (CSCRC), Australia

mojtaba.shahin@monash.edu, ali.babar@adelaide.edu.au, aufeef.chauhan@adelaide.edu.au

## Absract

Modelling and Simulation as a Service (MSaaS) is a promising approach to deploy and execute Modelling and Simulation (M&S) applications quickly and on-demand. An appropriate software architecture is essential to deliver quality M&S applications following the MSaaS concept to a wide range of users. This study aims to characterize the state-of-the-art MSaaS architectures by conducting a systematic review of 31 papers published from 2010 to 2018. Our findings reveal that MSaaS applications are mainly designed using layered architecture style, followed by service-oriented architecture, component-based architecture, and pluggable component-based architecture. We also found that interoperability and deployability have the greatest importance in the architecture of MSaaS applications. In addition, our study indicates that the current MSaaS architectures do not meet the critical user requirements of modern M&S applications appropriately. Based on our results, we recommend that there is a need for more effort and research to (1) design the user interfaces that enable users to build and configure simulation models with minimum effort and limited domain knowledge, (2) provide mechanisms to improve the deployability of M&S applications, and (3) gain a deep insight into how M&S applications should be architected to respond to the emerging user requirements in the military domain.

## 1. Introduction

Modelling and Simulation (M&S) are used for analyzing the behaviors of large-scale complex systems to understand and address certain types of issues such as performance and scalability before implementing and operating real systems [1]. While a simulation model represents a simplified version of a system of interest (i.e., some information might be missed), it enables one (e.g., a decision-maker) to gain an appropriate understanding of a system through providing a desired or required level of abstraction of a system [2, 3]. Over the last few years, many methodologies, processes, and associated tools such as AutoDEVS [4], M&S Life Cycle [5], and Component-based M&S Development [6], have been developed to support the development and deployment of M&S applications. According to Wang [7], these methodologies have costly and complex development and maintenance phases, require different people with different domain knowledge in each phase, may not be compatible with new web technologies (e.g., cloud), and may not truly be applicable in practice. On the other hand, integrating existing modelling and simulation tools in a simulation environment to build an integrated simulation platform is a challenging task [8].

In order to maximize the potential benefits of M&S applications, Modelling and Simulation as a Service (MSaaS) has emerged as a new paradigm to develop and deliver M&S applications following the as-a-service model of Cloud Service Provider (CSP) or Service-Oriented Computing (SOC) [9-11]. There are a number of similarities between MSaaS and Software as a Service (SaaS) [12, 13]. They both provide utilities to the end-users based on a pay-per-use model and aim at the rapid provisioning of services. However, MSaaS can provide support for the customization of simulations and deploying custom simulations. In this case, it can be closer to Platform as a Service (PaaS) [13]. Figure 1 shows the general architecture of MSaaS [12]. In this paradigm, a simulation service ranges from a small plug-in service to a full simulation system [14]. The multi-tenancy characteristic of SaaS allows multiple tenants can access a platform [15, 16]. This characteristic also guarantees that their private data, services, and system configurations are manageable according to each tenant's needs. The data is protected from being exposed to other tenants. Multi-tenancy for MSaaS requires the identification of tenants, filtering their data, and configuring services at different levels. The first stage is the tenants' authentication so that they can be authorized to access the desired resources. Several factors are to be considered for managing access, including user management, role management, permission management, resource management, and constraint management. Once authentication is performed, the next step is to analyze incoming traffic of a specific tenant so that the corresponding services can be mapped to the message router. The service registration center can be used to keep records of which models and simulations are assigned to a specific tenant. Finally, segregation among the user data is also provided at the data storage level.

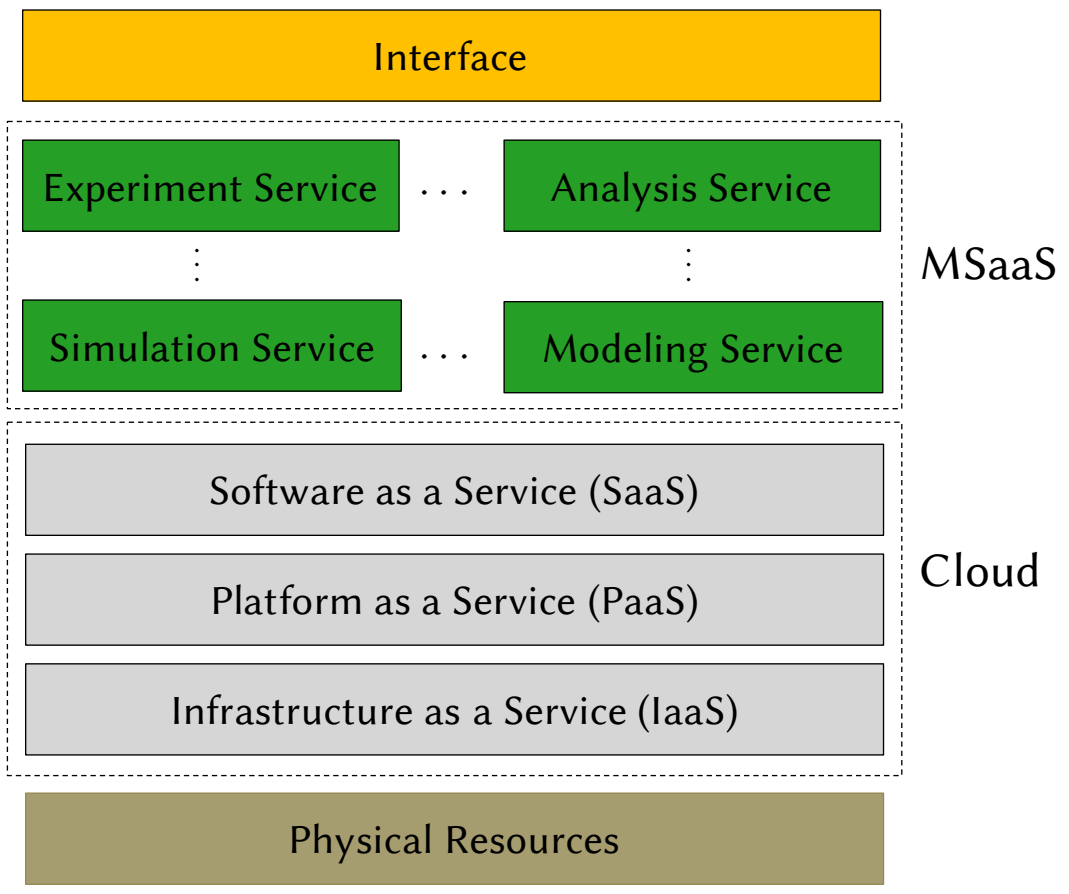

Figure 1. The general architecture of MSaaS

The MSaaS paradigm is expected to (1) provide facilities to develop scalable and composable simulations that can be deployed and executed in an on-demand fashion [9] and be available efficiently and cost-effectively [10, 17]. (2) It makes the underlying infrastructures, platforms, and the complexity of software transparent to users (e.g., developers and end users) by providing an intuitive user interface [12]. (3) It supports the rapid integration of multiple simulated systems and capabilities into a unified simulation environment [17, 18]. (4) It significantly reduces the need for human resources (e.g., domain experts) in configuring and running simulation models. (5) Finally, the MSaaS paradigm helps the users smoothly migrate their legacy simulation systems to a cloud environment. The MSaaS paradigm might also be associated with several risks, such as increased dependency on service providers and network connections and the need for new mechanisms to tackle security and privacy in distributed environments [17].

On the other hand, emerging MSaaS has the potential to drastically change the architectures of software systems and the infrastructures and platforms upon which software systems operate [12, 19]. Like for any other software-intensive system, software architecture is considered a backbone of MSaaS in order to deliver quality M&S applications. It is recognized that designing and evaluating architectures of large-scale software-intensive systems; specifically, those supporting or enabling MSaaS, is a complex and knowledge-intensive process [20, 21]. Designers or architects are expected to devise new architectural principles and strategies or evolve old ones to match the specific needs of systems hosting and delivering MSaaS. For example, a new class of questions such as *what should be provided as a simulation service*, *how correctly simulation services need to be composed for simulation experiments,* or *how to specify the boundaries of a simulation environment in a larger SOC environment* should be addressed in MSaaS architectures [22-24]. Architectural design decisions are made by utilizing technical (such as patterns, tactics, and quality attribute analysis models) as well as contextual, also called Design Rationale (DR), (such as design options considered, trade-offs made, assumptions, and design reasoning) knowledge [25]. Architects leverage both types of knowledge, also called Architecture Design Space (ADS), for identifying, assessing, and selecting suitable design options for design decisions and reasoning about the discarded and selected design options [26, 27].

While a growing body of the literature has explored different aspects of MSaaS, only two secondary studies have been conducted in this regard [12, 19]. Cayirci [12] conducted a survey on different aspects of MSaaS in 2013. He introduces MSaaS as a special form of Software as a Service (SaaS) and defines three types of MSaaS: modelling as a service, model as a service, and simulation as a service. Cayirci [12] also classifies MSaaS architectures based on deployment strategies into four groups: "standalone MSaaS applications", "federated standalone MSaaS applications", "composed MSaaS", and "automatically composed MSaaS". Cayirci elaborates that apart from security challenges (e.g., single point of failure) that any cloud service may face, simulation services leveraging cloud computing for building simulation experiments must deal with a new class of challenge called Multi-Level Security (MLS). MLS aims at ensuring the users with different security clearances can access a cloud-based simulation service via a secure flow control. The survey by Bocciarelli et al. [19] identifies four types of M&S approaches in Business Process (BP) domain: *conceptual*, *agent-based*, *DEVS-based,* and *discrete event*. The study [19] also reveals that M&S approaches in the BP domain have six open issues, such as they have low flexibility and customizability and require a business analyst with simulation skills.

Our observation is that there has been no significant effort aimed at building a body of knowledge about designing and evaluating softwarized infrastructures for MSaaS. There has been little reporting about the design processes and practices and the kinds of knowledge required or generated while designing and evaluating MSaaS in the previous secondary studies [12, 19]. This apparent lack of attention to a highly critical aspect of designing and evolving large-scale softwarized infrastructures for MSaaS presents a unique opportunity to understand and support knowledge-intensive processes and practices of software design in a very important domain. In this paper, we try to explore ADS related problems of softwarized infrastructures for MSaaS, through a Systematic Literature Review (SLR) [28]. This SLR aims to build appropriate guidance and/or effective supportive infrastructure for designing, evaluating, and evolving softwarized infrastructures for MSaaS. ADS can be developed and operated by capturing and managing the details of architectural challenges and solutions for MSaaS.

To the best of our knowledge, no secondary study has investigated ADS for MSaaS. This study is expected to provide a comprehensive and holistic overview of the state-of-the-art of architecture-related research on MSaaS for practitioners and researchers. (1) We discuss different cutting-edge architectures proposed in the literature for MSaaS from architectural drivers and quality attributes perspectives (**RQ1**). Moreover, we elaborate on the strengths and weaknesses of each architecture. We then compare and contrast the identified architectures with each other by using the requirements of an industrial, military M&S application. (2) We identify architectural solutions used to support or enable MSaaS (**RQ2**). We believe this big picture of the architecture aspect of MSaaS will be the key enabling factor for the development and deployment of M&S applications following the MSaaS concept.

Our findings reveal that (1) layered architecture style is the dominant architectural style to design MSaaS applications, followed by service-oriented architecture, component-based architecture, and pluggable component-based architecture; (2) interoperability, deployability, cost, performance, scalability, and configurability are the top six quality attributes that are of great importance in the architecture of MSaaS applications; (3) user interface design is one of the key divers of architecting process of MSaaS applications, by which end user should be able to build and configure simulation models with minimum effort and limited domain knowledge; (4) the MSaaS architectures proposed in the literature are not able to meet the critical user requirements of modern M&S applications appropriately.

Our contributions are (1) characterizing MSaaS architectures, (2) identifying a set of architectural solutions for MSaaS and mapping them to a well-known, M&S application development framework (i.e., Data Farming framework), (3) articulating the strengths and weaknesses of the-state-of-the-art architectures for MSaaS, and (4) identifying some areas for future research by highlighting the characteristics, limitations, and open issues of this research topic.

In the remainder of this paper, Section 2 introduces the design of our SLR. Section 3 presents the findings. In Section 4, we discuss threats to the validity of our review. Section 5 draws conclusions and discusses our main observations.

## 2. Research Method

We conducted an SLR by following the general guidelines proposed by Kitchenham and Charters [28]. The main goal of this SLR is to characterize the MSaaS architectures proposed in the literature. We formulated two research questions (RQs), which serve as the basis for this SLR. Table 1 shows the research questions, along with their motivations. In the following subsections, we describe the main steps of this SLR.

Table 1. Research questions and their motivations.

| Research Question | Motivation |
|---|---|
| **RQ1.** What are the-state-of-the-art architectures for MSaaS? | The answer to this research question provides an in-depth understanding of characteristics of the architectures proposed in the literature for MSaaS: the architectural styles employed by each architecture, the main drivers behind them, the limitations of each architecture, and the quality attributes are improved or satisfied by the proposed architectures. We also evaluate each architecture in the context of a modern M&S application in the military domain. |
| **RQ2.** What are architectural solutions proposed in the literature for MSaaS? | The goal of this question is to gain a deep insight into architectural solutions (e.g., architectural practices, patterns, techniques, architectural decisions) that support or enable the implementation of the MSaaS paradigm. |

### 2.1 Search Strategy

We decided only to use the Scopus search engine to find potentially relevant papers. Our previous experiences show that Scopus provides the most comprehensive search engine for retrieving research papers in computer science as it indexes a large number of journals, conferences, and workshops [29, 30]. Other research [31] also shares that running a well-designed search string on the Scopus search engine can retrieve the overwhelming majority of the relevant papers indexed by other search engines, including IEEE Xplore, ACM Digital Library, SpringerLink, Wiley Online Library and ScienceDirect. We designed a search string and ran it on Scopus. We instructed the Scopus search engine to search only in the papers' title, abstract, and keywords. It should be noted that the search string was incrementally finalized after conducting several pilot searches and verifying whether it could have identified the well-known primary studies in the MSaaS paradigm.

***TITLE-ABS-KEY*** *((“simulation-as-a-service” OR “simulation-as-service” OR “simulation as a service” OR “simulation as service” OR “model*-as-a-service” OR “model*-as-service” OR “model* as a service” OR “model* as service” OR “model* and simulation-as-a-service” OR “model* and simulation-as-service” OR “model* and simulation as a service” OR “model* and simulation as service” OR “simulation and model*-as-a-service” OR “simulation and model*-as-service” OR “simulation and model* as a service” OR “simulation and model* as service” OR “model* & simulation-as-a-service” OR “model* & simulation-as-service” OR “model* & simulation as a service” OR “model* & simulation as service” OR “simulation & model*-as-a-service” OR “simulation & model*-as-service” OR “simulation & model* as a service” OR “simulation & model* as service” OR “model* and simulation application as a service” OR “model* and simulation application as service” OR “model* & simulation application as a service” OR “model* & simulation application as service” OR “simulation and model* application as a service” OR “simulation and model* application as service” OR “simulation & model* application as a service” OR “simulation & model* application as service” OR “MSaaS” OR “M&SaaS” OR “SMaaS” OR “S&MaaS” OR “MaaS”) AND (“architect*” OR “design*”))*

## 2.2 Study Selection

The execution of the search string on Scopus in August 2019 retrieved 352 papers. We filtered the retrieved papers by applying a set of inclusion and exclusion criteria (See Table 2). Specifically, we included the studies proposing an architecture or architectural solutions for MSaaS. We excluded the papers that only focused on mathematical techniques for simulation models or performing the evaluation of a MSaaS-based application without describing its architecture and architectural solutions. We conducted the study selection process in three steps. In each step, we excluded irrelevant papers, and the papers selected in each step were used as input for the next step. In the first step, we made the decision about the inclusion or exclusion of a paper by reading its title and keywords. In the second round, we applied the inclusion and exclusion criteria to the papers' abstract. In the last step, we read the full detail of the selected papers. A paper had to satisfy all inclusion criteria to be considered as a primary study for this SLR. The study selection process resulted in 31 primary studies, which were used for data analysis in this SLR.

Table 2. Inclusion and exclusion criteria

| **Inclusion Criteria** | |
|---|---|
| **I1** | A study that is peer-reviewed and available in full-text. |
| **I2** | A study that reports an architecture or architectural solutions for MSaaS. |
| **Exclusion Criteria** | |
| **E1** | Non-peer-reviewed papers such as editorials, position papers, keynotes, reviews, tutorial summaries, and panel discussions. |
| **E2** | Non-English papers. |
| **E3** | A study reports mathematical techniques for simulation models or evaluates a MSaaS-based application without describing its architecture and architectural solutions. |

## 2.3 Data Extraction

Information extraction from the selected primary studies was conducted based on the research questions and documented in an Excel spreadsheet file. Apart from demographic data (e.g., publishing date), we wrote a critical summary of each proposed architecture for MSaaS and collected the type of used architectural style, the limitations of the proposed architectures, problem(s) addressed by the reported architectural solutions, and the quality attributes improved/addressed to answer RQ1. To answer RQ2, we extracted the best practices, lessons learned, decisions, patterns, techniques, and challenges reported at the architecture level for MSaaS.

## 2.4 Data Analysis

### 2.4.1 Analyzing RQ1

The extracted data for RQ1 were analyzed using qualitative methods, including open coding and constant comparison [32]. We first performed open coding iteratively in parallel with extracting information from the reviewed papers. This step captured architectural styles, main drivers, quality attributes, possible limitations (weaknesses), and addressed architectural problems in each selected primary study and assigned a label (i.e., code) to each. Second, we applied constant comparison to compare the codes identified in one selected primary study with the codes that emerged from other primary studies. The last step iteratively grouped these emergent codes to generate higher levels of categories.

Apart from individually articulating the weaknesses and strengths of each identified architecture, we aimed at comparing and contrasting the identified architectures with each other. Given practitioners and relevant stakeholders (e.g., policymakers) using evidence-based practice are one of the main audiences of SLRs [28, 33], we used the high-level user requirements of an industrial, modern M&S application in the military domain. We leveraged those requirements as a criteria-based evaluation framework to assess the state-of-the-art MSaaS architectures (See Tables 6, 9, 12, and 15). While M&S is rapidly being applied to many domains (e.g., transportation), the selection of the military domain in this study can be justified by two facts. Firstly, M&S has been widely used in military organizations for decades [17], wherein M&S products are considered as high-assets. Secondly, the concept of MSaaS, along with software architectures supporting MSaaS, appeared as a response to evolving critical needs (e.g., cost) in the military domain [10, 18]. We chose Strategic Response 2 (SR2) project, *Simulation for Future Operating Concept Development*, as a modern M&S application in the military domain [34]. SR2 is part of a five-year research initiative program called Modelling Complex Warfighting (MCW) Strategic Research Initiative (SRI), which has been launched in 2017 by Defence Science and Technology (DST) Group[1] in Australia [34]. SR2 aims at developing innovative simulations based on the MSaaS concept for analyzing complex joint operational concept exploration. SR2 is also to develop a High-Performance Computing (HPC)-based infrastructure for executing experiments of the modeled force against simulated threats. To that end, SR2 needs to provide mechanisms to address the following key requirements: support wide range types of simulation models, enable parallel execution of closed-loop simulations, build loosely coupled simulations, use Data Farming to execute experiments and efficiently manage the data artifacts produced within Data Farming. The architecture of such Data Farming infrastructure should be compatible with the DST Group classified network or with DST Group's strategic computing infrastructure plans. A list of high-level user requirements of SR2 is presented in Table 3. It should be noted that SR2's requirements might not be a comprehensive list of all requirements of MSaaS-based applications. For example, Bocciarelli et al. [19] propose six high-level requirements that the next generation of MSaaS architectures in the BP domain should address. These requirements range from supporting and provisioning different simulation services and modelling languages, to building modelling services automatically, to being compatible with model-driven standards and technologies. Other relevant requirements, such as service interoperability, service composability, and deployment automation support, could also be considered to evaluate the identified MSaaS architectures. Such requirements are mostly covered as part of articulating each identified architecture's weaknesses and strengths (See Tables 5, 8, 11, and 14).

Table 3. High-level user requirements of SR2 as a modern M&S application in the military domain

| Req# | Description |
|---|---|
| **R1** | Support diverse types of simulations, e.g., simulations written in different programming languages, built by different simulation builders. |
| **R2** | Simulations should be executed independently. |
| **R3** | Simulations should not pass information directly to one another (e.g., they need to use a messaging queue). |
| **R4** | Enable visually monitoring and optimizing the behavior of simulations during their execution (e.g., reformulating or modifying a long batch of simulation runs). |
| **R5** | Support parallel execution of simulations. |
| **R6** | Support convenient and intuitive interfaces. |
| **R7** | Treat simulation as black-box, i.e., simulation runs should be transparent to users. |
| **R8** | Simulation jobs may need to be scheduled, load balanced and prioritized. |

### 2.4.2 Analyzing RQ2

Similar to RQ1, we used open coding and constant comparison to analyze the extracted data for RQ2 (i.e., the architectural solutions proposed in the literature for MSaaS), eventually generating eight core categories of the architectural solutions (See Section 3.2). We further mapped the core categories of the architectural solutions (i.e., Design Knowledge) to the five activities of Data Farming framework (Figure 2). The rationale behind this mapping is that both the architecting process and Data Farming framework highlight the importance of decision making. As discussed in the Introduction section, the (design) decisions made during architecting are considered as the first-class entity, and the architect is referred to as a decision-maker (i.e., making the right decisions among many alternatives at the right time) [35]. Data Farming framework is a promising simulation-based analysis approach to support decision making in complex systems [36, 37]. Data Farming framework was developed by the North Atlantic Treaty Organization (NATO) Science and Technology Organization [38]. Data Farming has become a well-known framework, which uses a collaborative, iterative, and question-centric process to combine the main activities of the M&S applications development process including "Rapid Scenario Prototyping", "Model Development", "Design of Experiments", "High-Performance Computing", and "Analysis and Visualization". Data Farming helps decision-makers (e.g., architect) to gain a holistic view of M&S applications development and enables them to effectively answer operational questions, e.g., "what-if" questions that traditional M&S processes are not able to address [39]. It can help decision-makers at different levels, ranging from low-level tactical decisions, to architecture design decisions (e.g., what configuration should be chosen for a communications network), to strategic decisions [36].

[1] www.dst.defence.gov.au

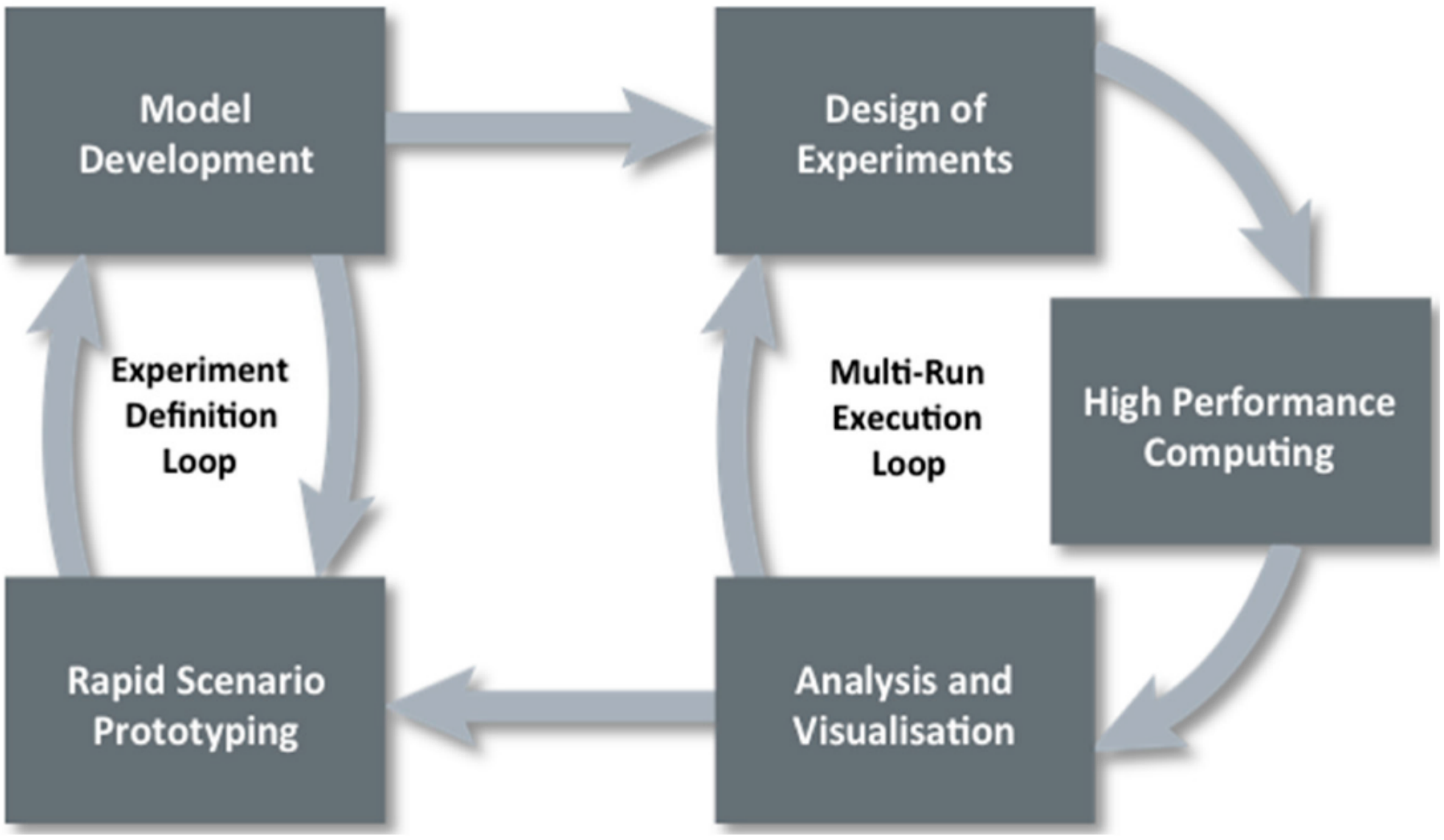

Figure 2. Data farming "loop of loops" – taken from [38]

## 2.5 Selected Primary Studies

Figure 3(a) illustrates the number of papers published per year. The 31 papers were published between 2010 to August 2019, reaching its peak in 2017 (i.e., 41.9% published in 2017). As illustrated in Figure 3(b), out of 31 primary studies, 12 (39%) papers were published in journals, and 19 (61%) in conferences and workshops. Table 19 in the Appendix provides a detailed view of the 31 papers and their venues. The quality of the selected papers is measured by the number of citations. We obtained the citation counts from Google Scholar on 15 March 2020. Table 19 shows that the number of citations ranges from 0 to 41, with an average of 10.9. Considering 80.6% of the studies have been published between 2016 and 2018, this low number is justifiable. As shown in Table 19, we used a set of indicators to measure the quality of venues, including SCImago Journal Rank (SJR)[2], CORE Ranking[3], impact factor, h5-index[4], and h-median[5]. 12 journal papers were published in 10 distinct journals, in which 8 are Q1 journals, and 2 are Q2 journals. 5 out of 12 journal papers were published in journals with the impact factor higher than 4, 4 papers come from journals having an impact factor between 2 and 4, and 3 papers are from journals without impact factor. The average h5-index for all venues is 37.73, in which the IEEE Access journal with the h5-index of 89 has the highest h5-index.

[2] https://www.scimagojr.com/journalrank.php
[3] http://www.core.edu.au/
[4] https://scholar.google.com/intl/en/scholar/metrics.html#metrics
[5] https://scholar.google.com/intl/en/scholar/metrics.html#metrics

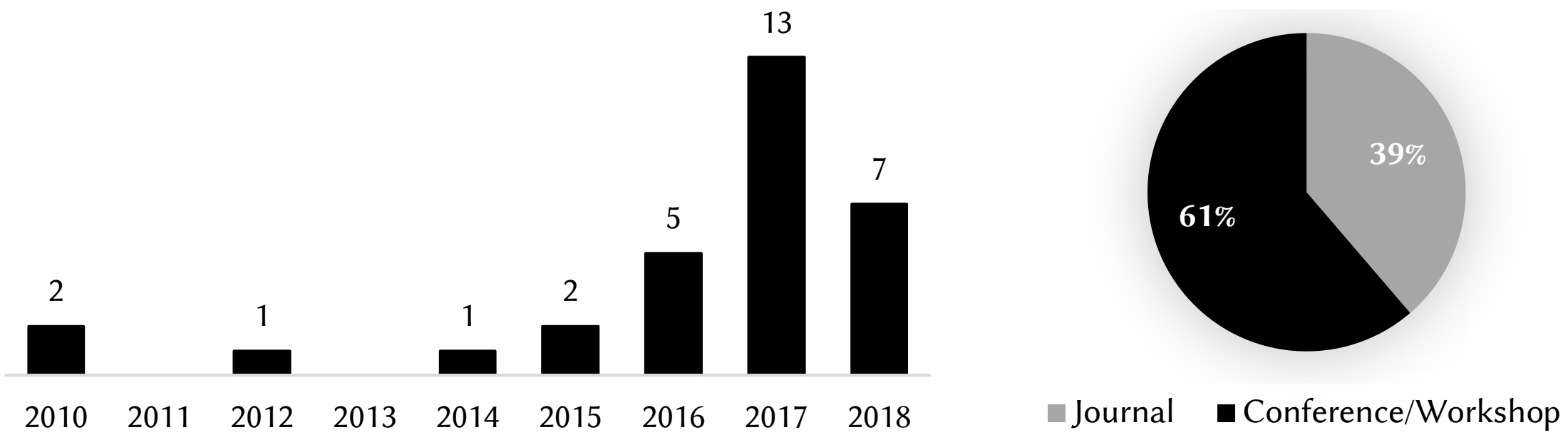

Figure 3. (a) Distribution of primary studies by publication year. (b) Distribution of primary studies by venue

# 3. Results

## 3.1 Architectural Styles, Drivers and Quality Attributes (RQ1)

In this section, we report the architectures that have been introduced in the literature for MSaaS, along with the architectural styles, main drivers behind the chosen architectures, and their potential impacts in terms of quality attributes. At the highest level, the identified architectures are grouped based on four architectural styles, including layered architecture, component architecture, pluggable component-based architecture, and service-oriented architecture. Architectural style describes a set of design rules to define a vocabulary of components and connectors, and local or global constraints, which are used to structure a system and support interconnections within and between components [40]. It is worth mentioning that a system's architecture may be a combination of different architectural styles [41]. For instance, a SOA architecture composing different services might be organized into a layered architecture.

The architectures in each style are further classified by using the drivers behind them (See Tables 4, 7, 10, and 13) and the quality attributes that they address or improve (See Tables 5, 8, 11, and 14). We also show which layer, component, service, or technique in the identified architectures contributes to improve or address quality attributes. In total, we found six architectural drivers (i.e., **D1** to **D6**), which significantly influence the structure of and decisions made for MSaaS architectures.

Finally, we articulate the strengths and weaknesses of each of the identified architectures and then assess the state-of-the-art MSaaS architectures by using the critical requirements of SR project discussed in Section 2.4.1 (See Tables 6, 9, 12, and 15). A list of high-level user requirements of SR2 is presented in Table 3.

### 3.1.1 Layered Architecture

Layered architecture is the most common architectural style used in MSaaS. In this style, the primary element is the layer, and the functionalities of a system are hierarchically grouped into separate layers, in which each layer needs to have a well-defined interface and provide operations to other layers [41].

Developing M&S applications with web services and cloud technologies may present challenges such as difficulties in the integration process of complex M&S applications and a need for domain knowledge from end users to perform the integration process [42]. The main reason for such a requirement is that this class of systems includes heterogeneous data and services. Wainer and Wang [42] propose a Mashup Architecture for Modelling and Simulation (MAMS) with five layers: Cloud, Box, Tag Ontology, Wiring, Mashup Application layers. MAMS would leverage mashup technology to tackle the challenges associated with developing and integrating M&S applications. In this architecture, M&S resources (e.g., model) are considered as boxes and are seamlessly integrated as mashups. ***Impacts***: Leveraging mashup and RESTful web services make integrating and deploying processes of M&S services much easier.

Hüning et al. [43] present a MSaaS solution called Multi-Agent Research and Simulation System (MARS) to handle the complexity of large-scale agent-based simulation models. The architecture of MARS leverages layered architectural style, which has been chosen because layered architecture is compatible with the organization's GIS files structure. Each layer contains a distinct type of agent or environmental data [44]. The complexity of the simulation models is handled in the back-end, while the domain experts are provided a user-friendly web interface (i.e., called MARS Websuite) to manage their data and models. ***Impacts***: Agent Shadowing technique and Docker-based virtualization (described in Section 3.2) used in MARS have positive consequences on the scalability and deployability of simulation models respectively.

The ability to easily configure remote simulation environments and running simulation environments more efficiently (e.g., in parallel) are the main drivers to adopt a MSaaS approach leveraging cloud computing facilities, as reported by Harri and colleagues [45]. The architecture proposed in [45] is organized into two layers: front-end and back-end layers. While the front-end layer includes a web service/server to enable domain experts to configure remote simulations easily, the back-end layer provides facilities to conduct remote simulations through a controller (i.e., simulation engine) and High Capacity Computing (HCC) platform. ***Impacts***: This architecture is expected to ease the configuration of remote simulation models.

Building simulation environments manually can be time consuming and error-prone [46]. Liu et al. [46] introduce a cloud-based simulation framework employing a layered structure to tackle these issues. The proposed architecture provides customized MSaaS to users in a flexible and automatic way and includes a user interface layer, resource manager layer, VM pool layer, physical layer. ***Impacts***: From a user's perspective, the introduced architecture brings higher levels of manageability (e.g., easy configuration of simulation models) and flexibility (e.g., customizing service requirements). Furthermore, it improves the reusability and deployability of the customized simulation environments.

Cayirci et al. [47] report hTEC JMSOS services and architecture that make M&S support available through simulation as service for effectively integrating space capabilities into joint military operations. The hTEC architecture [47] has a layered architecture, in which the layers' structure is heavily influenced by cloud service models, including MSaaS. With this architecture, MSaaS is represented as a special form of Software as a Service (SaaS) and is structured into three different layers: model/service layer (i.e., providing models as service), modelling/service composition layer (i.e., creating a mashup from the models produced by the model layer) and simulation/session layer (i.e., executing the models composed in the composition layer). It should be noted that a security service is implemented as a sublayer in the model/service layer. ***Impacts***: The proposed architecture is expected to improve the interoperability and composability of simulation services.

The layered architecture proposed in [48] utilizes and enhances the Internet of Things (IoT) concepts to describe a network of *simulated things* (i.e., Internet of Simulation – IoS). The simulated things might be simulation models, tools, and systems from different domains and applications. Apart from leveraging core layers in a cloud (e.g., PaaS), IoS includes Simulation as a Service (SIMaaS) and Workflow as a Service (WFaaS) layers. SIMaaS layer extends the SaaS layer hosted in a cloud to expose individual simulations to the wider IoS. Users interact with IoS through WFaaS layer, which provides a mechanism to aggregate the individual simulations provided by SIMaaS and build large-scale co-simulation systems. WFaaS leads to hiding the details of the SIMaaS and Cloud platform from users. ***Impacts***: IoS is expected to facilitate the integration and deployment of simulations.

Recently, SaaS concept has gained significant attention as an effective and efficient paradigm of providing models (e.g., numerical models) as service [49, 50]. Chen et al. [49] introduce a new architecture to describe the concept of Model as a Service (MaaS). Their architecture has four layers, including physical machines layer, virtual machines, numerical models layer, and user layer. The physical processes such as *irrigation* and *river seepage* in a groundwater system are provided to this architecture as a service, which can be simulated by numerical models. ***Impacts***: Exposing numerical models with MaaS gives flexibility to users to build their own customized models based on specific needs.

Suram et al. [50] propose an architecture to present engineering models as microservices. While the architecture proposed in [50] is organized into several layers, the model space layer leverages the microservices approach to break down a monolithic engineering model into stateless and independent models. It is because adopting a monolith approach for engineering models is challenging as these models may be developed by different teams having different domain knowledge. The models may be implemented using different programming languages by following different modelling practices. ***Impacts***: With this architecture, the resilience of an engineering system is enhanced because models do not retain any state information (i.e., no single point of failure). Furthermore, having a collection of independent models would increase the scalability of the whole system and enable developers to reuse these models in building complex system models.

Another study [2] propose a MSaaS platform named *SOASim* to support web-based modelling services. The components of this platform are grouped into two distinct layers: front-end layer and back-end layer. The front-end layer provides several interfaces that allow one to manipulate and store models and specify the tasks required to perform a simulation-based analysis. The back-end layer includes two sub-layers: the server layer consisting of components for managing simulation activities and cloud-based infrastructure layer handling the execution and deployment of simulation components into cloud-based infrastructures. ***Impacts***: SOASim is adaptable to several application domains and modelling languages.

Bocciarelli et al. [51] propose a platform called BP-MSaaS to simplify M&S-based Business Process (BP) analysis. In the existing M&S-based BP analysis approaches, BP models do not (semantically) correlate with simulation models. To address this limitation, the platform utilizes the MSaaS concept, containerization technology, and model-driven approach. The architecture of the platform is structured into three layers: *web layer*, *middleware layer*, and *cloud infrastructure layer*. Simulation as complementary to BP modelling is realized in *middleware layer* by defining four distinct services: *modelling services* to create, edit and store BP models; *transformation services* to automatically generate the respective executable simulation models; *simulation services* to deploy and execute simulations in the cloud, and *presentation services* to visualize simulation results. ***Impacts***: Whilst leveraging the MSaaS paradigm is expected to improve the interoperability and composability of BP simulation services, automatically generated modelling services using a model-driven approach reduces the efforts required for managing (e.g., executing) BP simulation models. Additionally, simulation services are packed and deployed into self-contained Docker images to improve their deployability.

The users (e.g., a researcher or technician) of debris-flow simulations need to be provided all their required functionalities into a single environment as they can easily perform their tasks [52]. Rosatti et al. [52] develop a partially-open web service ecosystem for this purpose, which is an all-in-one solution to manage (e.g., produce) debris-flow simulations. This class of simulations is used for assessing debris-flow hazards. The ecosystem called WEEZARD (WEbgis modElling and haZard Assessment for mountain flows: an integRated system in the clouD) leverages layered architecture approach to organize its decoupled components. The components communicate with each other via well-defined, but not presently standard interfaces. In this architecture, the application layer can be hosted by any client machine as it has low computational overhead and provides interfaces to manipulate simulation jobs and results. The middleware layer as the core layer of WEEZARD is responsible for handling security, converting data, and transferring data between layers/components. The bottom layer (i.e., data layer) deals with storing and retrieving data to/from the file system or an object-relational database (e.g., PostgreSQL). ***Impacts***: WEEZARD enables users to produce and analyze simulation jobs at a lower cost and in less time.

Modelling in geoscience (e.g., climate model) is associated with unique challenges [53]. For example, model setup is a complex process, huge (scalable) computing resources are required to simulate models, and model simulation can generate substantial amounts of data. Li et al. [53] propose a Model as a Service (MaaS) framework built on IaaS to address these challenges. Their framework employs a four-layer architecture containing a cloud computing platform layer, geoscience model image repository layer, MaaS middleware layer, and MaaS user layer. From the implementation point of view, the architecture includes four main software elements: a *web portal* for uploading model inputs, requesting a model run and receiving model outputs; a virtual agent called *MaaS server* to monitor model runs and to automatically provision computing resources (See *parallelizing ensemble model run* and *concurrent user requests* mechanisms in Section 3.2.4.2); a *MaaS engine* on the cloud to compile and execute a model into its own specific ready-to-go-environment (i.e., model VM); and a *data server* to manage model outputs in a centralized database. ***Impacts***: Apart from improved interoperability using the ready-to-go model environment, the proposed architecture automatically provisions computing recourses to have lower cost and better performance.

Carillo et al. [54] develop a SIMulation-as-a-Service (SIMaaS) infrastructure to enable non-experts users to configure and run distributed simulations on a cloud smoothly. The architecture of SIMaaS consists of three layers. At the bottom layer, the necessary infrastructures are provisioned to run simulations by Amazon EC2. SIMaaS utilizes a customized version of StarCluster[6] toolkit on top of its infrastructure layer to simplify managing and configuring Amazon EC2 instances. StarCluster is customized via plug-ins. On the top of the customized StarCluster, D-Mason framework exists, which provides the distribution of agent-based simulations at the framework level [55]. This helps such distribution to be transparent to end users. The architecture of the D-Mason framework is based on master/workers pattern, in which the master node splits a large-scale simulation job into smaller partitions (workers). The partitions are run independently from each other and communicate using Apache ActiveMQ[7], a message broker. Interface layer includes a decoupled web interface system that exposes its functionalities via web services. ***Impacts***: The proposed architecture not only eases the configuration process of computing resources (e.g., Amazon EC2 instances) through the customized StarCluster but also has a positive impact on performance and cost.

Taylor et al. [56] propose a CloudSME Simulation Platform (CSSP) to migrate a range of legacy commercial simulation applications to cloud computing and expose them as cloud-based simulation services. Whilst improving interoperability among simulation services is one of the main goals of CSSP, the most important reason behind the development of CSSP is that users can obtain results from the simulation runs in a shorter time. CSSP is composed of *simulation application layer*, *CSSP layer*, and *cloud resource layer*. *Simulation application layer* can host both web-based designs and non-web designs of simulation applications. *CSSP layer* leverages *gUSE* (Grid User Support Environment) and CloudBroker platforms. *gUSE* platform enables users to conveniently specify and run the workflow of simulation tasks on a Distributed Computing Infrastructures (DCI). ***Impacts***: Enhanced security, interoperability, and response time are the main impacts of the proposed platform.

Ibrahim et al. [57] provide a framework named SIM-Cumulus for large scale network simulations. Execution time and energy cost are the main targeted quality attributes. Performance is assessed using large scale wireless network simulations. The framework supports automation of the required virtual machine instance creation and parallel simulation executions. The framework is supported by a multilayer architecture consisting of a virtual infrastructure layer, cloud instance management layer, virtual platform layer, cloud infrastructure management layer, and system accessibility layer. ***Impacts***: Reduction in carbon footprint and cost of simulation to support sustainable simulation services.

Similar to the work of Taylor et al. [56], Liu et al. [24] develop a platform called Cloud-based computer Simulation (CSim) to operate simulation software tools from cloud computing infrastructure. *Users*, *web browser*, and *simulation cloud* constitute the layers of CSim's architecture. *Simulation cloud* itself has two layers: SIMulation as a Service (SIMaaS) and Virtualized Computing Environment (VCE). SIMaaS provides the functionalities of a simulation tool to users in the form of web services such as modelling as a service, execution as a service, and analysis as a service. The major module in this architecture is Cloud Manager Module (CMM), which controls all the services in SIMaaS layer and the resources in VCE layer. VCE is a pool of physical and virtual resources that can be accessed and manipulated by APIs. ***Impacts***: CSim enables users to easily access simulation techniques.

[6]http://star.mit.edu/cluster/
[7]http://activemq.apache.org/

Table 4. The main drivers behind the MSaaS architectures that are designed based on the layered architecture style

| Architectural Driver | Key Points and Included Papers | # |
|---|---|---|
| **D1**. Simulation models should be built and used by users with no or minimum domain knowledge. | ▪ Integrating M&S applications should not require extensive domain knowledge and expertise [42].<br>▪ The complexity of the back-end simulation system should be transparent to domain experts [43].<br>▪ Users with a low computational background should be able to set up and run simulations in the cloud easily [54]. | 3 |
| **D2**. Simulation models should be built, integrated, and configured in a simple (e.g., low effort) and automatic way. | ▪ Integrating M&S applications should be simple [42].<br>▪ Simulation environments should be built automatically [46].<br>▪ Simulation models should be easily configured [45].<br>▪ Integrating different types of simulations should be automatic [48].<br>▪ The users (e.g., researchers) of models should be able to build customized models for different purposes instead of building numerical models from scratch [49].<br>▪ Business analysts should perform M&S-based BP analysis in an effective and effortless manner [51].<br>▪ All debris-flow simulation jobs should be produced, performed, and analyzed in *one single environment* [52].<br>▪ Models in geospatial science should be easily set up [53].<br>▪ Resources (e.g., computing resources) in geospatial science should be managed automatically [53].<br>▪ Users should be able to easily use simulation techniques with low cost [24]. | 9 |
| **D3**. Simulation models should be compatible with other standards, services, and etc. | ▪ The architecture design should be compatible with the GIS structure [43].<br>▪ Space capabilities need to be integrated into joint military operations by making M&S support available [47].<br>▪ MSaaS platform should be compatible with innovative web-based modelling services [2]. | 3 |
| **D4**. Simulation models should be run in parallel. | ▪ Users should be able to run parallel executions using virtual machine instances [57].<br>▪ Simulations should be run in parallel [45].<br>▪ Users should receive the results of simulation runs fast [56]. | 3 |
| **D5**. Simulation models should be decoupled from other others. | ▪ Engineering models should be decoupled from each other, in which model developers should have autonomy in choosing the structure of and programming languages for the engineering models [50]. | 1 |

Table 5. The quality attributes improved/addressed by the MSaaS architectures that are designed based on the layered architecture style and their corresponding layers/components

| Quality Attributes | System Layer or Component | Key Points and Included Paper |
|---|---|---|
| Interoperability | RESTful web services and Tag Ontologies | ▪ Ontologies and platform-neutral APIs support integration [42] |
| | Service Composition Layer | ▪ Creates mashup from the models [47] |
| | WFaaS layer | ▪ Supports aggregation of individual simulations to build large scale co-simulation systems [48] |
| | Containerization and model-driven approach | ▪ Semantically correlates simulation models [51] |
| | MaaS engine | ▪ Compiles and executes ready-to-go model environments [53] |
| | gUSE (Grid User Support Environment) | ▪ Supports specification and running the workflow simulation tasks on distributed infrastructures [56] |
| Deployability | Cloud Box | ▪ Models are treated as boxes and can be easily deployed [42] |
| | MARS Websuite | ▪ User-friendly web interfaces for managing data and deploying models [43] |
| | VM Pool Layer and Physical Layer | ▪ Automation of deployment process [46] |
| | SIMaaS | ▪ SIMaaS utilizes PaaS and IaaS to deploy services on the underlying cloud infrastructure [48] |
| | Simulation Services and Cloud Infrastructure Layer | ▪ Simulation services utilize the cloud infrastructure layer to facilitate deployment [51] |
| Cost | MaaS Server | ▪ Monitors and automatically provisions computing resources [53] |
| | Master/Worker Pattern | ▪ Master node splits large simulation jobs into small worker partitions. The partitions can run independently, having a positive impact on cost [54] |
| | Cloud Manager Module | ▪ Manages simulation server and cloud infrastructure resources cost-effectively [24] |
| | Cloud Infrastructure Management Layer | ▪ Reduces the energy cost of running the simulations [57] |
| Configurability | Mashup Application Layer | ▪ Mashup technology to handle development [42] |
| | Front-end Web Service | ▪ Enables domain experts to configure remote simulations [45] |
| | Resource Management Layer | ▪ Automatically builds and configures simulation environments [45] |
| | StarCluster | ▪ Simplifies management and configuration of Amazon EC2 instances [54] |
| Performance | MaaS Server | ▪ Monitors execution of models and automatically provision required computing resources [53] |
| | D-Mason Framework | ▪ Transparent distributions of agent-based simulations on underlying infrastructure [54] |
| | gUse (Grid User Support Environment) | ▪ Supports specification and running workflow of simulation of workflow tasks on distributed computing infrastructure [56] |
| Reusability | Model Space Layer | ▪ Breakdowns monolithic engineering model into stateless and independent models to support reusability [50] |
| | Containerization Technology and Transformation Services | ▪ Models are transformed into respective executable simulation models and deployed on the cloud [51] |
| Scalability | MARS Websuite | ▪ Separation of concerns in different layers so that each layer handles a distinct type of data [43] |
| | Model Space Layer | ▪ Microservice approach to breakdown a monolithic engineering model into multiple stateless independent models [50] |
| Efficiency | Resource Manager Layer | ▪ Automatically manages virtualized and physical resources to run simulations [46] |
| | Cloud Manager Module (CMM) | ▪ Efficiently manages simulation services and resources in the virtualized computing environment [24] |
| Security | CSSP Layer | ▪ Uses grid user support environment and cloud CloudBroker platforms to isolate resources [56] |
| Customizability | Resource Manager Layer | ▪ Provides flexibility for customization and automatically managing virtualized and physical resources to run simulations [46] |
| | Numerical Models Layer and User Layer | ▪ Provides flexibility to users to customize and build their own models [49] |
| Composability | Modelling/Service Composition Layer | ▪ Creates mashup of simulation models [47] |
| | Modelling Services | ▪ Facilitates creating, editing and simulation services [51] |
| Usability | User Layer | ▪ Facilitates users to build their own customized models [49] |
| | Web Browser | ▪ Allows access, model, execute and analyze simulations [24] |
| Modularity | StarCluster | ▪ Simplifies managing and configuring infrastructure resources to run distributed simulations on the cloud [54] |
| Sustainability | Cloud Infrastructure Management Layer | ▪ Reduces execution time and energy cost by automating virtual machine instance creation and parallel simulation executions [57] |
| Resilience | Model Space Layer | ▪ Enhances the resilience of the system by using stateless microservices as simulation models does not retain any state information [50] |
| Adaptability | Backend Server Layer | ▪ Includes components to adapt and manage simulations activities [2] |
| Ease of Development | Mashup Application Layer | ▪ Mashup and RESTful web services allows the implementation of simulation logic into multiple languages and support seamless integration [42] |

Table 6. Mapping high-level user requirements of SR2 to the MSaaS architectures that are designed based on the layered architecture style.

| **Ref** | **High-level Requirements of SR2** | | | | | | | | **Strengths** | **Weaknesses** |
|---|---|---|---|---|---|---|---|---|---|---|
| | **R1** | **R2** | **R3** | **R4** | **R5** | **R6** | **R7** | **R8** | | |
| MAMS [42] | √ | - | - | - | - | √ | √ | - | ▪ Requires no domain knowledge.<br>▪ Integrates and executes diverse M&S resources produced by different tools. | ▪ Allows continuous improvement and evolution *only after* a simulation finish.<br>▪ Composing M&S resources based on event-driven architecture imposes a sequential composition. |
| MARS [43] | - | - | √ | √ | - | √ | √ | - | ▪ Allows automatic dependency injection among dependent layers.<br>▪ Allows users to optimize simulation models during simulation runs (i.e., providing real-time visualization).<br>▪ Enables users to perform advanced analytics. | ▪ Affected the most by changes in GIS data structure.<br>▪ No support for running multiple simulations in parallel. |
| [45] | - | √ | - | - | √ | √ | √ | - | ▪ Has the ability to support different virtualization techniques (Xen and KVM). | ▪ The use of a central mechanism, called *controller*, to manage (e.g., distribute) simulations on HPC may create a single point of failure. |
| [46] | - | √ | - | - | - | √ | √ | - | ▪ No need to manually configure simulation environment (e.g., support automatic resources configuration).<br>▪ Provides two types of interfaces: GUI and command-line interface. | ▪ No support for parallelization.<br>▪ May impose virtualization overhead and service overhead. |
| hTEC [47] | √ | √ | - | - | - | √ | √ | - | ▪ Supports a wide range of simulation types.<br>▪ Consideration of security requirements at the architecture level, not as an afterthought.<br>▪ Service composition process does not pose a time constraint. | ▪ Lacks in providing a mechanism for reformulating simulation models during execution.<br>▪ Requires complete information to standardize a service properly.<br>▪ Needs standard convention if a service following different standards.<br>▪ Requires identifying the time-sensitive part of a simulation service at *design time.* |
| IoS [48] | √ | √ | - | - | - | √ | √ | - | ▪ Enables a wide range of simulations to work together in a distributed co-simulation environment.<br>▪ Provides workflows as services and distinguishes them from simulation services. | ▪ Requires more attention to usability, interoperability, and dependability |
| [49] | √ | √ | - | - | - | √ | √ | - | ▪ Support different kinds of numerical models. | ▪ Requires numerical models to be created *offline* with an *extensive knowledge domain.*<br>▪ Lacks in providing parallel executions of numerical models. |
| [50] | √ | √ | √ | - | - | √ | √ | - | ▪ Enables integrations of simulation models written in multiple programming languages with different internal structures.<br>▪ No state needs to be kept by simulation services (e.g., easy scalable).<br>▪ Models are decoupled from the rest of the system. | ▪ Needs to invoke engineering models in a sequential manner.<br>▪ No support for software and hardware virtualization |
| | | | | | | | | | ▪ Resilient architecture (no single point of failure). | |
| SOASim [2] | √ | √ | - | - | - | √ | √ | - | ▪ Requires minimal/no effort to build simulation models (i.e., use a model-driven approach).<br>▪ Provides relevant simulation engine per each simulation model. | ▪ Requires changing or extending the proposed model-driven approach to manage new simulation model types.<br>▪ Storing simulation models in XML format has overhead. |
| BP-MSaaS [51] | √ | √ | - | - | - | √ | √ | - | ▪ Complements BP lifecycle with simulation.<br>▪ Consideration of security requirements at the architecture level, not as an afterthought.<br>▪ Requires minimal/no effort to build simulation models (i.e., use a model-driven approach).<br>▪ Each simulation service, along with its dependents (e.g., runtime), is considered as an executable image in a Docker container. | ▪ Requires adding new simulation engines to manage non-BP application domains. |
| WEEZARD [52] | - | √ | - | √ | √ | √ | √ | √ | ▪ Consideration of security requirements at the architecture level, not as an afterthought.<br>▪ Enables users to monitor and analyze the behavior of simulation jobs at runtime. | ▪ Only supports one type of numerical model (i.e., TRENT2D model).<br>▪ Only a certain number of simulation jobs can run concurrently.<br>▪ Is not portable (i.e., numerical model written in Fortran 90 is compiled with processor-specific optimization in a Windows OS).<br>▪ Requires extensive collaboration between different groups with different skills. |
| [53] | √ | √ | - | - | √ | √ | √ | √ | ▪ Requires minimum effort to set up complex geoscience models.<br>▪ Supports the ensemble run of a model with different configurations (i.e., parallelization at the experiment level).<br>▪ Enables to respond to multiple run requests of a model at the same time.<br>▪ Enables online visualization and analytics tools to use model outputs.<br>▪ Supports scheduling model requests. | ▪ Publishing a new simulation model into a MaaS platform is a complicated process (e.g., requires extensive collaborations between the modelers/researchers and MaaS providers).<br>▪ Requires a large computing pool to truly implement *parallelizing ensemble model run* and *concurrent user requests* mechanisms.<br>▪ Heavily relies on users to upload model input and configuration files manually. |
| SIMaaS [54] | - | √ | √ | √ | - | √ | √ | - | ▪ Using Master/Workers pattern in D-Mason (i.e., as core simulation framework) enables execution of long-running simulation jobs. | ▪ Heavily depends on user inputs (e.g., simulation's parameters).<br>▪ Only supports one type of simulation model (i.e., agent-based simulation). |
| CloudSME Simulation Platform [56] | - | √ | - | - | √ | √ | √ | √ | ▪ Consideration of security requirements at the architecture level, not as an afterthought. CSSP uses the security mechanisms provided by CloudBroker platform in a transparent way.<br>▪ Supports parallel and distributed simulations.<br>▪ Enables migration of different applications from different software vendors to the cloud. | ▪ Lacks the use of HPC on the cloud.<br>▪ Lacks semantic information to consolidate the different software [58]. |
| CSim [24] | - | √ | - | - | √ | √ | √ | √ | ▪ Execute simulation tasks at the *function granularity* rather than the *class (entity) granularity* (i.e., it distributes the functions of a simulation task in various nodes/clouds). | ▪ No support for optimizing the behavior of simulations at runtime. |
| SIM-Cumulus [57] | - | - | - | - | √ | - | - | - | ▪ Enables parallel execution of simulation activities | ▪ Supports only network-based simulations. |

### 3.1.2 Component-based Architecture

With component-based architecture style, a system is decomposed into a set of reusable components that expose well-defined interfaces [41]. The component-based architecture style has been used in 4 of the reviewed studies [9, 59-61].

van den Berg et al. [9] have leveraged containerization as an alternative approach to virtual machines in the simulation applications built on High Level Architecture (HLA). HLA is a component-based architecture for distributed simulation environments. ***Impacts***: Whilst it is argued that containerization is more suitable for small M&S applications rather than monolithic M&S applications, the enhanced HLA-based simulations with containerization would have simpler development and deployment process and higher scalability.

Fischer et al. [59] argue that conventional simulation tools (e.g., AnyLogic[8]) need to be enhanced for dynamic and service-oriented environments. They propose a generic approach so-called *Simulation Environment* to provide simulations tools as service. The component-based architecture of SE aims at exposing as much as possible the functionalities of a set of simulation tools into independent-tool functionalities. Hence, the functionalities of a simulation are divided into two parts: *Generic Functionalities* and *Simulation-tool Specific Functionalities*. The approach automatically generates and executes a simulation model and supports two types of simulations: discrete event and agent-based simulations. The main component of this architecture is *coordination module*, which interprets simulation requests and determines the workflow execution of other components. ***Impacts***: Improved interoperability among simulation tools and reusability are the main consequences of the proposed architecture.

Analyzing and modelling large-scale scientific data can be a pain for scientists as they may not have in-depth programming experience and computational background. Evans and Nikolic [60] report a system, *Prometheus,* to expose PyRhO module (i.e., a computational platform for optogenetics) as Modelling as a Service (MaaS). The system leverages Docker technology for virtualization and application packaging as well as for smoothing the configuration of a data science environment. A Docker container contains PyRhO and its dependencies (e.g., NumPy). ***Impacts***: *Prometheus* enables computational models to be used and reproduced by scientists in a collaborative manner.

Table 7. The main drivers behind the MSaaS architectures that are designed based on component-based architecture

| Architectural Driver | Key Points and Included Papers | # |
|---|---|---|
| D2. Simulation models should be built, integrated, and configured in a simple (e.g., low effort) and automatic way. | ▪ Mental simulation service should be introduced to openEASE, a cloud engine, to support robots on how to handle their situation [61].<br>▪ Conventional simulation tools should be provided as a service to be used in dynamic and service-oriented environments [59].<br>▪ Scientists should be able to easily use and configure a data science environment [60]. | 3 |

[8]https://www.anylogic.com/

Table 8. The quality attributes improved/addressed by the MSaaS architectures that are designed based on component-based architecture and their corresponding layers/components

| Quality Attributes | System Layer or Component | Key Points and Included Paper |
|---|---|---|
| Reusability | Generic Functionalities Simulation | ▪ Automatically generates simulations by combining generic functionalities with simulation specific functionalities [59] |
| Deployability | Containerized Environment | ▪ High-level architecture based simulations with containerization enhance deployment process [9] |
| Cost | Containerized Environment | ▪ Containerization supports need-based deployment of the resources [9] |
| Collaborative Environment | PyRhO in docker containers | ▪ Enables collaborative usage of computational models [60] |
| Ease of Development | Containerized Environment | ▪ Containerization simplifies development [9] |
| Usability | Docker Technology | ▪ Used for virtualization and application packaging that supports reusability in terms of ease of deployment [60] |
| Reproducibility | PyRhO in Docker Containers | ▪ Enables collaborative reproduction of computational models [60] |
| Scalability | Containerized Environment | ▪ Containerization supports cost-effective scalability [9] |
| Interoperability | Component-based Simulation Environment | ▪ Improves interoperability among simulation tools by dividing simulation tools into independent tool functionalities [59] |

Table 9. Mapping high-level user requirements of SR2 to the MSaaS architectures that are designed based on component-based architecture.

| Ref | High-level Requirements of SR2 | | | | | | | | Strengths | Weaknesses |
|---|---|---|---|---|---|---|---|---|---|---|
| | R1 | R2 | R3 | R4 | R5 | R6 | R7 | R8 | | |
| [9] | √ | √ | - | - | - | - | √ | - | ▪ Requires minimal/no effort to run a distributed simulation.<br>▪ Simplifies the development process of simulation applications. | ▪ Containerizing monolithic simulation applications may pose key problems.<br>▪ Automatic container build is affected by dynamic command-line interfaces.<br>▪ Automatic container build is affected by dynamic GUIs.<br>▪ Containerized application needs to decouple GUI from its main application.<br>▪ Needs base images and standard command-line options to configure the application through environment variables [62]. |
| [59] | √ | √ | - | - | √ | - | √ | √ | ▪ Supports two types of simulation paradigms: discrete event and agent-based simulations.<br>▪ Requires minimal/no effort to generate and execute simulation models.<br>▪ Supports parallelization and synchronizes various simulation runs | ▪ Require changes to data models when introducing new simulation tools.<br>▪ Only supports AML format to exchange data.<br>▪ Difficult to use AML format for large data structure as the size of AML file increases. |
| Prometheus [60] | - | √ | - | - | - | √ | √ | - | ▪ Automatically deletes inactive resources. | ▪ Lacks mechanisms to prohibit non-registered users.<br>▪ No support for data persistence. |
| [61] | - | √ | √ | - | - | √ | √ | - | ▪ Provides rich reasoning techniques for analyzing the results of simulation experiments. | ▪ Heavily depends on Prolog queries. Users need to have some Prolog knowledge.<br>▪ No support to run multiple agent simulations in parallel. |

### 3.1.3 Pluggable Component-based Architecture

Pluggable component-based architecture style (also known as plug-and-play architecture [63]) has been reported in two papers [64, 65]. A pluggable component aims to extend the functionality of a core application [63, 66]. In a pluggable system, the core application is separated from the extensions (plug-ins) through a plug-in manager.

Existing real-world simulators such as SUMO (Simulation of Urban MObility), a transportation system simulator, for STEM education (Science, Technology, Engineering, and Mathematics) have the following issues: (i) it is hard to be used by non-domain experts (e.g., students); (ii) they cannot be accessed and shared remotely by multiple users; (iii) there is a lack of scalability and data persistence when there are multiple users [65]. Caglar et al. [65] propose a modelling and simulation framework, called $C^2$SuMo, integrated with pluggable components. Within the architecture of $C^2$SuMo, SUMO constitutes the core application. The objective of $C^2$SuMo is to hide the complexity of SUMO from high school students. This is achieved by the components (e.g., $C^2$SuMo middleware) that are implemented as plug-ins to SUMO. ***Impacts***: The proposed system provides a collaborative environment that can be used and comprehended by K-12 students. It improves scalability by employing multiple SUMO simulators in a cloud-based infrastructure.

The SIMaaS Cloud Middleware is another pluggable component-based architecture proposed in [64]. This system utilizes a Linux container-based virtualization solution instead of virtual machine technology. This decision is made to address the challenges (e.g., lack of having a real-time response) that may occur during the deployment of a specific type of simulation system in traditional cloud infrastructure (i.e., virtual machines). This type of simulation system is characterized by having the simulation model that either need multiple runs or produce different outcomes for various parameters. The key component of this system is *SIMaaS Manager*, which manages resources and handles user requests. The *SIMaaS Manager* component has a pluggable design, which allows a virtualization approach (full virtualization, paravirtualization, or lightweight containers) to be selected or switched dynamically. ***Impacts***: With this architecture, simulation models can be executed in parallel with lower cost and minimum resources.

Table 10. The main drivers behind the MSaaS architectures that are designed based on pluggable component-based architecture

| Architectural Driver | Key Points and Included Papers | # |
|---|---|---|
| **D1**. Simulation models should be built and used by users with no or minimum domain knowledge. | ▪ Simulation system should be used by non-domain experts [65]. | 1 |
| **D2**. Simulation models should be built, integrated, and configured in a simple (e.g., low effort) and automatic way. | ▪ Simulation system should provide a mechanism to aggregate the results from the simulation models with different outcomes [64]. | 1 |
| **D4**. Simulation models should be run in parallel. | ▪ Simulation models should run in parallel [64]. | 1 |
| **D6**. Simulation models should be built in a collaboration environment. | ▪ Simulation system should support collaboration and coordination among users [65].<br>▪ Simulation system should be scalable and support data persistence when there are multiple users [65]. | 1 |

Table 11. The quality attributes improved/addressed by the MSaaS architectures that are designed based on pluggable component-based architecture and their corresponding layers/components

| Quality Attributes | System Layer or Component | Key Points and Included Paper |
|---|---|---|
| Performance | Container-based Virtualization Solution | ▪ Supports multiple runs of simulations [64] |
| Scalability | SUMO Simulators | ▪ Provides scalability by employing multiple simulators in a cloud-based infrastructure [65] |
| Cost | SIMaaS Manager | ▪ Cost-effectively manages resources by considering user requests [64] |
| Data Persistence | C²SuMo Middleware | ▪ Supports data persistence for multiple users [65] |
| Efficiency | SIMaaS Manager | ▪ Has a pluggable design, which allows dynamic selection and switching of virtualization approach [64] |
| Collaborative Environment | C²SuMo Middleware | ▪ Hides complexity and supports the implementation of the components as plugins [65] |

Table 12. high-level user requirements of SR2 to the MSaaS architectures that are designed based pluggable component-based architecture

| Ref | High-level Requirements of SR2 | | | | | | | | Strengths | Weaknesses |
|---|---|---|---|---|---|---|---|---|---|---|
| | R1 | R2 | R3 | R4 | R5 | R6 | R7 | R8 | | |
| C²SuMo [65] | - | √ | - | - | √ | √ | √ | - | ▪ Enables parallelization and scalability by employing multiple simulators in the cloud.<br>▪ Requires no domain knowledge. | ▪ Is restricted to one simulation domain model (i.e., traffic simulation).<br>▪ All map data should be converted into a specific format.<br>▪ Needs an administrator to initially configure it. |
| [64] | √ | √ | √ | - | √ | √ | √ | √ | ▪ Has the ability to dynamically switch among different virtualization techniques.<br>▪ Support two distinct types of simulation models.<br>▪ Has the ability to intelligently schedule and prioritize simulation jobs. | ▪ Heavily depends on user inputs (e.g., simulation execution time should be determined in the design phase).<br>▪ May induce sluggish performance as it uses Shipyard as container manager. |

### 3.1.4 Service-oriented Architecture

Preisler et al. [1] have adopted a service-oriented architecture and approach to design and implement large-scale energy network simulations based on *simulation as a service* concept. With this approach, each distinct type of simulation (e.g., multi-agent simulation or 4GL model) is encapsulated by a Simulation Service Component utilizing REST web services. The proposed approach enables building large-scale, loosely coupled simulation systems, yet they can be integrated and invoked in a unified way. ***Impacts***: Improved interoperability and scalability of simulation systems participating in a co-simulation system are the main consequences of this approach.

Zheng et al. [67] have developed a WebGIS-based service architecture consisting of loosely coupled simulations services (modules) to support users in analyzing spatial configurations of functionally interrelated facilities. Two main services in this architecture are Simulation Service (a loosely coupled service) and Geographic Service (a stateless service). Whilst the first one provides mechanisms for scenario editing, simulation control, and monitoring, the latter one aims at evaluating installation plans through the comprehensive perception of human behaviors. ***Impacts***: Usability, availability, and extensibility are three quality attributes that are positively impacted by the proposed architecture.

In [68], the authors develop a Distributed Simulation Framework instantiated from VIntEL NATO Reference Architecture for an underwater glider simulator. The architecture of this framework follows the MSaaS concept by employing a service-oriented architecture style to organize and connect M&S services. RESTful web services are used to realize this architecture, in which simulation engine, models, and other services (e.g., map services) are provided as web services. Nevertheless, the simulation engine is conceded to HLA without any web service (i.e., direct connectivity). It is mainly because the high rate of HLA/REST conversion might negatively impact the performance of the final system. From the deployment perspective, the REST service architecture is organized into three subsystems: a REST front-end servlet hosted by a servlet container, a set of microservices written in different programming languages at the back-end, and security system. Apart from relying on HTTPs transport protocol to ensure the confidentiality and integrity of data on client side, three tactics are used to enhance security in the proposed architecture: firewall, Demilitarized Zone (DMS), and authenticating and authorizing users. DMS acts as a secure front-end line and is surrounded by two firewalls to control input and output traffics. Before redirecting users to back-end web services, the users are authenticated and authorized to access proper web services. ***Impacts***: Among other quality attributes (e.g., configurability and deployability), interoperability and security are the most improved by the proposed architecture.

Pax et al. [69] provide a platform for crowd simulation as a service. The platform allows multiple simulations to run simultaneously and joint operations over multiple entities. The system is built by following the principles of microservices architecture. Agent-based simulations services are supported by a simulation backend, which consists of a library of recorded simulations. ***Impacts***: The proposed architecture facilitates the accurate measurement of the sensors to run different types of simulations.

Bocciarelli et al. [70] present a platform that is based on model-driven architecture and cloud computing. Underlying cloud infrastructure is abstracted with a cloud interface and containerization layer. On top of it, there is an application server hosting services for: model repository, modelling service, transformation service, simulation service, and presentation service. Features of the platform are exposed to end users via platform portal. A cloud-based simulation repository consists of model transformations, simulations models, system models, and executable code. The platform is extended using a layered architecture, which consists of multiple components [71]. Simulation service manager orchestrates actual and simulated services with the help of a business process model and business process manager engine. The extended platform is built using service-oriented architecture and cloud computing. ***Impacts***: The platform facilitates the service-in-the-loop approach by combining real and simulated services for simulating business process tasks.

Drawert et al. [72] develop an integrated environment, called StochSS (Stochastic Simulation as a Service), for simulation and analysis of biochemical models. The main goal of the StochSS is to provide a collaborative environment that the biologists with basic computer skills can execute and scale out their simulations in a cloud infrastructure with a click-of-a-button. The modelling and simulation process in StochSS includes four activities. *Problem specification* is to build models and/or convert them to the desired format using an easy-to-use model editor. *Simulation manager* selects one of four built-in simulation tools to execute a simulation model based on the simulation type. *Model analysis* includes a set of tools to support parameter estimation and parameter sensitivity. Finally, the goal of *output and visualization* activity is to provide an interactive 3D visualization of biomedical systems and simulation results. ***Impacts***: The StochSS facilitates collaboration among biologists, and thanks to the MOLNs cloud platform described in Section 3.2.4.3, it enables convenient and efficient execution of simulations.

Table 13. The main drivers behind the MSaaS architectures that are designed based on service-oriented architecture

| **Architectural Drivers** | **Key Points and Included Papers** | **#** |
|---|---|---|
| **D1**. Simulation models should be built and used by users with no or minimum domain knowledge. | ▪ Non-expert users should be able to easily create and scale up a model [72]. | 1 |
| **D2**. Simulation models should be built, integrated, and configured in a simple (e.g., low effort) and automatic way. | ▪ Different types of simulation systems should be integrated in a unified way [1].<br>▪ Simulation models should be extensible (i.e., easily being added, changed, deleted) [67].<br>▪ Model-driven architecture support for simulation design [70].<br>▪ Orchestration of actual and simulated services with business process model [71]. | 3 |
| **D3**. Simulation models should be compatible with other standards, services, etc. | ▪ The architecture of the simulation engine should allow enforcing the security requirements required for NATO security accreditation [68]. | 1 |
| **D4**. Simulation models should be run in parallel. | ▪ Parallel execution of multiple simulations [69]. | 1 |

Table 14. The quality attributes improved/addressed by the MSaaS architectures that are designed based on service-oriented architecture and corresponding services/components

| Quality Attributes | Service Type or System Component | Key Points and Included Paper |
|---|---|---|
| Interoperability | REST Web Services | ▪ Supports loosely coupled simulation systems [1] |
| | Back-end Server Layer | ▪ A set of microservices written in different languages supports interoperability |
| Deployability | Multi-layer Architecture | ▪ System is decoupled into three subsystems and deployment of each subsystem can be managed independently [68] |
| | Automatic Deployment | ▪ Executes and scales out simulations in a cloud infrastructure with the click of a button [72] |
| Autonomy | Cloud Interface and Containerization | ▪ Abstracts underlying cloud infrastructure for autonomous resource allocation [70] |
| | Simulation Service Manager | ▪ Abstracts services and simulation business processes for autonomous simulation provisioning [71] |
| Collaborative Environment | Model Editor | ▪ Collaboratively builds and converts models to the desired format using an easy to use model editor services and an interactive 3D visualization system [72] |
| Scalability | Multi-agent Simulation | ▪ Encapsulates simulation service components to facilitate deployment of RESTful simulation [1] |
| Usability | Stateless Geographic Service | ▪ Supports users for in analyzing the functionality of interrelated services [67] |
| Extensibility | Loosely Coupled Simulation Service | ▪ Supports extensibility by providing provision for scenario editing, simulation control and monitoring [67] |
| Portability | REST Services | ▪ REST services written in multiple programming languages can easily be ported on different types of hosting environments [68] |
| Efficiency | Model Analysis | ▪ Set of tools to support efficient parameter estimation and parameter sensitivity [72] |
| Performance | Cloud Interface and Containerization | ▪ Dynamic acquisition of the underlying infrastructure resources |
| Availability | Stateless Services | ▪ Supports easy replication of the services for enhanced availability [67] |
| Security | Security System | ▪ HTTPs transport protocol, firewall, demilitarized zone, and authentication and authorization contributes towards achieving security [68] |

Table 15. Mapping high-level user requirements of SR2 to the MSaaS architectures that are designed based on service-oriented architecture

| Ref | High-level Requirements of SR2 | | | | | | | | Strengths | Weaknesses |
|---|---|---|---|---|---|---|---|---|---|---|
| | R1 | R2 | R3 | R4 | R5 | R6 | R7 | R8 | | |
| [1] | √ | √ | - | - | √ | √ | √ | - | ▪ Supports and integrates different simulation types and models in a co-simulation environment.<br>▪ Enables parallel execution of simulation services. | ▪ Not very scalable to complex, distributed multi-agent system based simulations as the size of the system increases, it becomes very hard to encapsulate it with a *Simulation Service Component*. |
| [67] | - | √ | √ | √ | - | √ | √ | - | ▪ Implementation of simulation service as a loosely coupled resource allows other resources to use it in any other place.<br>▪ Enables users to receive real-time simulation results.<br>▪ Implements geoprocessing services as stateless services. | ▪ No support for parallelization.<br>▪ Only supports agent-based models. |
| [68] | √ | √ | - | √ | - | √ | √ | - | ▪ Consideration of security requirements at the architecture level, not as an afterthought.<br>▪ Supports different types of models. | ▪ Deviates IEEE Standard for M&S HLA because the simulation engine is directly connected to HLA (i.e., connection to HLA should be via a web service). |
| StochSS [72] | √ | √ | - | - | √ | √ | √ | - | ▪ Automatically transforms one type of biomedical model (e.g., well-mixed concentration model) to another type (e.g., well-mixed discrete model).<br>▪ Enables external analytics tools to leverage simulation results.<br>▪ Requires no more than basic computational knowledge to manage and deploy biological models. | ▪ MOLNs' functionalities (e.g., parameter sweep) are accessible only via a Command Line Interface (CLI), not as a Web UI. Hence, users need to learn commands. |
| [69] | - | - | - | √ | √ | - | √ | - | ▪ Allows multiple simulations to run in parallel.<br>▪ Supports joint operations over multiple simulation entities. | ▪ The use of older simulation data from a library of simulation records can produce inaccurate results. |
| [70] | √ | - | - | - | - | √ | - | - | ▪ Provides model-driven architecture support. | ▪ Interfaces of the simulation are exposed only via platform portal, not as web services. |
| [71] | √ | - | - | - | - | √ | - | - | ▪ Supports business process modelling.<br>▪ Supports combining real services with simulated services. | ▪ Results of real services can impact the results of simulation (when simulation is run as a combination of simulated and real services). |

### 3.1.5 Cross-Analysis of Architectural Styles

Table 16 shows which quality attributes are addressed or improved by each architectural style, along with the number of studies assigned to each style and quality attribute. The following patterns can be driven from Table 16: (1) The majority of quality attributes (14 out 24) are only reported in one study per architectural style. This pattern mainly happens in the component-based architecture style and pluggable component-based architecture style, as there are no dominant quality attributes in these styles. This may imply that the improved/addressed quality attributes can be more attributed to individual applications (e.g., a specific component or a service is responsible for the improvement) rather than the given style. (2) 15 studies out of 31 studies employing the layered architecture report that interoperability (6 studies), deployability (5 studies), and cost (4 studies) are effectively addressed. We argue that the layered architecture style plays a role in this scenario. (3) SOA style also positively influences deployability, autonomy, and interoperability because two studies report each of them. Moreover, autonomy is only reported by studies employing SOA style.

Table 16. Number of studies assigned to each quality attribute and architectural style

| QA | Layered | Component | Pluggable | SOA | QA | Layered | Component | Pluggable | SOA |
|---|---|---|---|---|---|---|---|---|---|
| Interoperability | 6 | 1 | - | 2 | Collaborative Environment | - | 1 | 1 | 1 |
| Deployability | 5 | 1 | - | 2 | Reproducibility | - | 1 | - | - |
| Cost | 4 | 1 | 1 | - | Data Persistence | - | - | 1 | - |
| Configurability | 2 | - | - | - | Autonomy | - | - | - | 2 |
| Performance | 3 | - | 1 | 1 | Extensibility | - | - | - | 1 |
| Reusability | 2 | 1 | - | - | Portability | - | - | - | 1 |
| Scalability | 2 | 1 | 1 | 1 | Usability | 2 | 1 | - | 1 |
| Efficiency | 2 | - | 1 | 1 | Modularity | 1 | - | - | - |
| Security | 1 | - | - | 1 | Sustainability | 1 | - | - | - |
| Customizability | 1 | - | - | - | Resilience | 1 | - | - | - |
| Composability | 1 | - | - | - | Adaptability | 1 | - | - | - |
| Availability | - | - | - | 1 | Ease of Development | 1 | 1 | - | - |

## 3.2 Mapping Architectural Solutions to Data Farming "Loop of Loops" (RQ2)

This section presents and classifies the architectural solutions identified in the literature for MSaaS (RQ2). Moreover, the architectural solutions (i.e., Design Knowledge) are mapped into the five activities of the Data Farming framework.

### 3.2.1 Rapid Scenario Prototyping

Rapid Scenario Prototyping (RSP) and model development, as two first steps of Data Farming framework, work closely with each other in the experiment definition loop in Figure 2. RSP formulates the specifics of a scenario to be simulated [73]. These specifics include model requirements for model development, the measurements that simulation needs to collect, and required input and output data for simulation.

Our review has identified a set of studies that aims at developing convenient and intuitive interfaces with which users can smoothly manage and manipulate simulations data and models, run simulation, and evaluate results. For example, MARS [43] is deployed in production as two separate parts: MARS Websuite and simulation system. MARS Websuite provides a web interface for modelers and model developers to configure a simulation system. In another example [59], the provided user interface enables users to configure simulation systems online or offline. In another study [60], the intuitive interface is achieved by running PyRhO GUI in a browser-based Jupyter Notebook interface. This interface is lightweight and interactive and is composed of familiar widgets (e.g., sliders). In another study [54], Carillo et al. suggest that the intuitive interface has been achieved by leveraging Google material design guidelines.

In [65], $C^2$SuMo architecture hides the complexity of SUMO from end users via an interface as K-12 students can use and comprehend it. The interface is composed of a Google Map interface, a Google Map API, and a mashup (i.e., created by integrating Google Charts). The reason to use Google Map API is that Google Map interface is static and does not support executable semantics. Furthermore, the functionalities of $C^2$SuMo are exposed via a low-level HTTP-based API, with which users (e.g., developers) and client applications can manipulate (e.g., query) simulations and their underlying infrastructures.

The work reported in [68] leverages a REST front-end interface to provide functionalities to end users, such as managing simulation instances and manipulating simulation status and parameters. The interface is decoupled from the simulation engine and runs as an independent process. This makes the interface to be an optional entity in the proposed architecture. This function is particularly useful for those applications where the simulation engine plays the role of a computation engine, as they do not require an interface. Similarly, the interface designed for SIMaaS infrastructure [54] is fully decoupled from the simulation part by exposing its functionalities via web services.

In [72], the designed interface also provides mechanisms to automatically convert one type of biomedical model (e.g., well-mixed concentration model) to another type (e.g., well-mixed discrete model). However, for converting the models with custom propensities, a user is needed to be involved in this process. In addition, the StochSS infrastructure [72] provides a Command Line Interface (CLI) for developers and administrators, which allows them to manage (e.g., deploy) StochSS instances as SaaS in a cloud IaaS (e.g., OpenStack).

### 3.2.2 Model Development

The second step aims at developing a simulation model that captures the basics of a scenario [73]. The results of a simulation model's runs and decision-makers' input can be leveraged to improve the quality of the simulation model iteratively.

#### *3.2.2.1 Simulation Model Characteristics*

We have found a set of papers aiming at characterizing simulation models and services [1, 24, 42, 48, 50, 53, 59, 67]. Our analysis found three characteristics of simulation models.

**Being stateless**: The authors in [1, 50] emphasize that stateless simulation models are more suitable for simulation systems as these simulation models can be seamlessly integrated. Suram and colleagues have proposed an architecture [50] in which models and solvers are designed stateless to be loosely decoupled from the rest of the system. As per the proposed architecture, the solver is a set of models that solve specific engineering problems, e.g., an equation. Hence, the stateless models are not allowed to retain or persist any state information; preferably, after a model execution, the execution workflow determines the following service to receive the state information. A *queuing system* has been implemented to transfer messages (i.e., the state information) between the stateless models at run time.

**Separating core assets from custom assets**: It is argued in [53] that handling geoscience models in the MaaS paradigm could be a complex task as each model might have its own heterogeneous input types and formats. To give an example, the number of the required configuration files as input for each model might be different. To address this challenge, the authors in [53] suggest that the model input should be abstracted into *model configuration* (e.g., geographic region) and *model data* and be prepared outside of MaaS server. Then, three standards channels can be defined for each model VM to locate *model data*, *model configuration*, and *model output*. With these three distinct channels, types and formats of models are encapsulated. Apart from improved interoperability, different configurations can be applied to each model execution without rebuilding the model environment. Liu et al. [24] propose a modelling approach to structure the simulated entity model in a cloud environment. With this modelling approach, a cloud-oriented simulation model is divided into the model description file, model configuration file, execution framework file, and computation file. Whilst a model configuration file specifies the desired format for the input, output, and interface of a model, the model configuration file captures the required configuration parameters for running simulations. On the other hand, both execution framework and computation files are designed as a dynamically linked library to store the implementation of the interfaces and realize the computation functions respectively. The rationale behind his decision is to enable model reusability at a finer-grained level: at the function level instead of the class level. This would be aligned with the goal of the E/C execution mode, which is described in Section 3.2.4.2.

Fischer et al. [59] emphasize that simulation model libraries (e.g., data conversion library) should be designed based on open-source standards. This enables them to be reused by more simulation tools. Furthermore, the basic/base part of a simulation model produced by a simulation tool is stored in a *simulation model component*. Then, each simulation tool can automatically generate use case specific simulation models by parametrizing the existing base simulation models.

Zheng et al. [67] also propose that facility (space) users' activity simulation in the proposed WebGIS based architecture should be designed and executed separately from other services. This loosely coupled service is run on the backend server and is considered as a reusable resource for other services. The approach is expected to improve the interactivity and availability of the whole simulation system. Liu et al. [24] highlight that the modelling services, analysis services, and experiment design services need to be independently interaction-intensive services. This helps stabilize and minimize the consumption of the resources, with which these services execute.

An approach focused on exposing *workflows* as service has been proposed by McKee et al. [48]. The authors indicate that apart from considering simulation as a service, the workflows need to be exposed as service (WFaaS). This helps abstract Simulation as Service and Cloud platform from users and to build a complex simulation system from individual simulations and execute it as a single virtual system.

**Providing functionalities via REST web services**: Wainer and Wang [42] propose an approach to exposing M&S resources, including simulation models, as RESTful web services using CloudRISE. The CloudRISE is a middleware, which extends RISE (RESTful Interoperability Simulation Environment) to support varied simulation models in a Cloud [74]. In this approach, M&S resources exchange information via HTTP methods (GET/PUT/POST/DELETE).

According to Preisler et al. [1], each type of simulation system (e.g., 4GL model and multi-agent simulation) should be encapsulated by a Service Component Simulation leveraging REST web services. This tactic results in simulations with different characteristics to be combined and invoked in a unified way. They also suggest that black-box simulations be built on IaaS layer in a cloud. An existing simulation application that has been implemented without cloud consideration is called black-box simulation. To provide a black-box simulation as a cloud-based service, the simulation as a service should be built on IaaS layer to abstract from physical resources rather than PaaS layer. It is because PaaS restricts black-box simulations to be compatible with its own APIs. Therefore, black-box simulations will not fully gain the anticipated benefits (i.e., scalability) from PaaS. Li et al. [53] also highlight that Model as a Service should be built on IaaS layer. They argue that with PaaS layer, users have no control over underlying resources, e.g., VM's operating system or storage/network. In contrast, it is suggested that a new or simple simulation should be built on PaaS layer as it can be encapsulated with a Service Simulation Component [1].

##### *3.2.2.2 Composing Simulation Models*

Whilst stateless and loosely coupled simulation services allow a fast and efficient integration process in simulation systems, the work in [42] describes an approach to find wirable (composable) M&S resources (e.g., simulation models). In this approach, a tag-tree knowledge base is developed to record the signature of each M&S resource (i.e., a M&S resource is called a *box* in MAMS [42]). A signature determines the inputs, outputs, and functions of a box. Second, a semantic approach is used to find appropriate boxes for the wiring process (i.e., composition process) based on their semantics. The signatures of the boxes are mined using tag-mining algorithms to add semantic (i.e., tags) to them. Then a domain-specific tag-tree ontology is built and learned based on tags. Finally, based on the existing tags and the learned tag-tree ontology, wirable resources to a given resource are suggested to users.

### 3.2.3 Design of Experiments

A simulation model may have several inputs or parameters, which can be manipulated to explore the simulation model's performance [73]. A mix of these inputs, called *design point*, constitutes a design experiment, which needs to be carefully chosen for running a simulation model.

##### *3.2.3.1 Store/Document Simulation Model Execution*

Storing the information about the execution of simulation models has been discussed in [43, 52, 53, 65, 68]. In [43], the relevant information about a simulation model such as the needed C# runtime, model code, GIS files, and configuration description are stored in a Docker container image. The main reason behind this decision is that the information stored in the Docker image can be later reused to execute the subsequent runs of a simulation model immediately. Additionally, the work in [43] documents the results of simulation model execution in CSV files. These files can later be utilized for advanced analytics with R. In another work, Caglar et al. [65] use MongoDB NoSQL database to store all user information (e.g., user interactions) and the results of simulation experiments. This architectural decision is made to have persistence data about user actions and past sessions. Similarly, in [53], model simulation outputs and related metadata are stored in a centralized database. Rosatti et al. [52] employ two storage approaches. Geospatial data is stored in PostgreSQL, which is extended by PostGIS to support spatial schema, but simulation information (e.g., simulation results) is directly stored in a file system. The work of [68] suggests that interoperability be highly achieved if simulation outputs are provided and stored based on open standards.

### 3.2.4 High-Performance Computing

High-Performance Computing (HPC) provides the required hardware (e.g., processors) and software (e.g., OS) to run a design experiment and to analyze and visualize the results of simulation runs [38].

#### *3.2.4.1 Lightweight Virtualization*

Using hypervisor-based virtualization techniques in the simulations that run in parallel is not suitable as they do not support real-time decisions (i.e., responding to user requests in a timely manner) [64]. This class of virtualization needs to boot the entire Operating System (OS) once a new virtual machine (VM) is scheduled. This delays the availability of new VMs. Shekhar et al. [64] leverage lightweight containers (e.g., Docker) to address this problem. Docker container virtualization technology outperforms hypervisor-based virtualization in this class of simulation as it has low overhead at runtime and improves resource sharing. In another attempt to use lightweight virtualization instead of hypervisor-based virtualization, Hüning et al. [43] report that having an easier deployment process and providing the same environment for development and production are the main reasons to use Docker for virtualizing simulation application. Yet, hardware virtualization is provided by hypervisor-based virtualization (i.e., Linux KVM). In [60], the reasons behind using Docker containers are reported as follows: Docker images can easily be created by a script and can be versioned. Docker also plays a key role in the architectures proposed in [51, 60] as a simulation service image, including the executable code, is deployed to a given Docker container to execute a simulation model.

#### *3.2.4.2 Resource Allocation*

The dynamic resource management algorithm proposed in [64] aims at parallelizing simulation jobs. The algorithm uses a Quality of Service (QoS)-based resource allocation policy to decide about the containers to be used for processing a simulation model. Time and cost are the two main factors in this decision. The algorithm ensures that the number of the assigned containers to each requested simulation model is minimum (i.e., cost), and simulation system responds to user requests within a reasonable amount of time. However, this algorithm heavily depends on the inputs (e.g., number of simulations, along with their parameters) provided by users.

van den Berg et al. [9] apply a Weave-based overlay networking technique to create a multi-host network, which allows inter-container communications across Docker hosts. With this technique, every container has its own unique ID address in an established overlay network. Therefore, a simulation component can be executed in a group of Docker hosts. Weave[9] technology is used to implement the *overlay networking* technique. Besides the *overlay networking* technique, the components such as Docker Compose, Docker Swarm, Consul, Registrator, and Weave Scope are used to automatically distribute and observe simulation components in a group of Docker hosts.

Orchestrating Docker containers in the proposed architecture by Evans and Nikolic [60] is performed by *tmpnb*[10]. *tmpnb* creates and assigns a Jupyter notebook server in a Docker container. Whilst both *tmpnb* and *JupyterHub*[11] could be used for this purpose, *tmpnb* has been chosen as it provides a better resource management mechanism. It is because, with *tmpnb*, inactive notebook servers can be automatically deleted after a period.

[9]https://www.weave.works/
[10]https://github.com/jupyter/tmpnb
[11]https://jupyterhub.readthedocs.io/en/latest//

In another reported architecture (so-called hTEC) [47], the session layer is designed to run the simulation models composed by its lower layer (i.e., modelling/service composition service layer). To run multiple instances of the composed services in the session layer, each instance needs to run with its own image of synthetic environment. With this technique, the original copy of a synthetic environment remains unchanged for subsequent usages. This technique not only allows users to run multiple instances of the composed services but also enables running each instance as a different type of simulation, e.g., time-stepped simulation vs. continuous simulation.

From the deployment perspective, hTEC is a distributed architecture that consists of a front-end and a back-end with several cloud-hosted services [47]. The delay between the back-end where the composed service runs and the front-end where the interactions with a system are performed can be a challenge in hTEC architecture. To address this challenge, a technique is employed to migrate those parts of a MSaaS that are time-sensitive (i.e., called cerebellum functions) to the machines close enough to the front end. These machines are expected to meet the delay constraints. The sensitive part of MSaaS should be designed in a way that can be separated from the rest of the services (i.e., designed as decomposable part). It is argued that this tactic also has a positive impact on security.

Caglar et al. [65] have modified a Python-based TraCI protocol to enable $C^2$SuMo simulation-as-a-service framework to communicate with multiple running simulations simultaneously. The original version of the TraCI protocol does not support communication with multiple running traffic simulations simultaneously. It is mainly because the default version of this protocol uses a singleton module-based approach, which is restricted to only one SUMO server running in a VM. One method to resolve this issue is to use a class-based approach instead of the singleton module-based approach [65]. With this method, the balanced webserver (i.e., Apache webserver) can handle each incoming request in its own SUMO instance, by which the instances of SUMO server can be spawned on-demand inside VMs. Additionally, hosting the backend SUMO simulation in a balanced web server helps manage the communication traffic between C2SuMo components.

In the architecture proposed in [52], a technique called "*asynchronous-mode process-request*" makes it possible for clients and servers to communicate. In this communication style, the client acts as an active entity and sends the "ExecuteRequest" message to a server. The server responds to this request by only providing a URL link (i.e., without pushing information). The decision is made because a simulation job might take time to complete, so users can check the status of a simulation job's execution at runtime. Since a server is able to respond to a limited number of user's requests at the same time, Rosatti et al. [52] develop a scheduling queue to schedule simulation jobs and assign a server to them.

In an experience report by Martinez-Salio [14], it is argued that combining sandbox and Certification Authority (CA) deployment models is expected to provide the most appropriate model for deploying simulation services in the MSaaS paradigm. Among other benefits (i.e., assuring the security of deployed services), this approach helps efficiently manage the resources utilized by simulation services during runtime. That means the internal services in a sandbox can monitor and check the resources accessed by the deployed simulation services. Not only are fewer resources needed, but the sandbox can also determine the maximum resources (e.g., memory) that a simulation service can use.

SOASim platform [2] provides different types of simulation engines for executing simulation models. The goal of this decision is to assign an appropriate simulation engine to each simulation model according to its characteristics. To give an example, performance-based business process simulation models are executed by eBPMN.

Li et al. [53] develop mechanisms to parallelize ensemble run of a model with different configurations and handle concurrent user requests for a model run. In both mechanisms, allocating model VMs to a model run request plays a significant role. In the parallelization mechanism, one specific model VM is automatically created for each model run in an ensemble run request. Whilst all model VMs utilize the same model VM image, the configuration of each model is different. In contrast, the mechanism for concurrently running a model uses different model VM images with different configurations for each model run request. Truly implementing both mechanisms necessitates a large computing pool; otherwise, a limited number of model runs can be served, and the pending model run requests can be started once a model VM finishes following the first-come-first-service policy. The architecture proposed in [53] significantly reduces the consumption of computing resources by immediately terminating VMs after a model run is finished.

CSim platform proposed in [24] adopts an Entity server/Calculation server (E/C)-based simulation execution mode to distribute and execute the simulation execution requests on different nodes in a processor pool. The simulated entities are deployed on entity servers. However, once the workload of the entity servers reaches a predefined level, the calculation servers are called to perform calculation tasks. The E/C mode dynamically migrates the simulated entities between the entity servers and calculation servers to have a dynamic load balancing during a simulation run.

A platform to allow the execution of multiple services simultaneously for operations over multiple entities is presented in [69]. Microservices architecture, combined with agent-based services, supports the accurate allocation of required resources for different types of simulations.

#### *3.2.4.3 Resource Configuration*

Our literature review has identified a set of papers [9, 45, 46, 53, 54, 64, 72] that provide mechanisms to easily configure resources (e.g., VM images) to run simulation systems. In [46], the authors introduce a *Virtual Machine Pool* layer in the proposed architecture. The virtualized infrastructures in this pool of resources can be manipulated from two aspects: VM image and VM entity. VM Monitor component in the proposed architecture is designed to manipulate VM images. VM monitor performs this task through maintaining an XML file, including image name, image location, valid time, OS name, and OS version. VMware Server as hypervisor is chosen to manage VM entities. Liu et al. [46] take a further step to configure network modules (e.g., VM image) in a virtualized environment. They apply a hash-map based method for this purpose. This method ensures that there is no conflict between the allocated IP addresses to VM images or entities. A hash function determines the IP address of a virtual machine, as it is highly unlikely that two users configure VMs in the same way.

Another kind of resource configuration mechanism concerns about selecting a virtualization approach at runtime [64]. In this paper, the *SIMaaS Manager* component in the proposed architecture adopts a pluggable design and employs the *strategy pattern* to determine a virtualization approach among full virtualization, paravirtualization, or lightweight virtualization approaches. In another work [54], the process of resources (e.g., Amazon EC2 instances) configuration is smoothed by customizing StarCluster toolkit with new plug-ins. The StarCluster plugin automatically builds a runnable and fully configured simulation environment.

van den Berg and colleagues [9] have reported the experiences of using Docker containers in developing and deploying HLA-based simulation systems. They have indicated that since users have no access to a container content, containerized applications should be configured by environment variables or via data containers. They also suggest that HLA-based simulation system should use a *standardized LRC (Local RTI Component) base image*, by which the configuration of the containerized application would be independent of specific HLA-RTI (i.e., RTI is a runtime infrastructure in HLA which provides mechanisms for federated interactions).

Harri and colleagues [45] have set up each VM as a fully configurable simulation environment. This is achieved by employing Kernel-based Virtual Machines (KVM), a virtualization approach for Linux on x86 and x86-64 processors. This technique enables configurable Windows and Linux images hosting simulation environments being executed on HCC (High Capacity Computing) platforms. This means that it is not restricted to a single HCC platform.

The work of [53] creates a ready-to-go environment (i.e., a model VM) for each new model to be published into the existing MaaS system. The pluggable-based design of MaaS architecture and the image-based mechanism provided by a cloud help implement this environment. In the first step, a model VM image is created for the newly published model by installing the model, along with dependent software libraries, into a “bare-metal” VM. Registering this new model VM image into MaaS is the next step, in which the positions of model input and output files on a model VM are determined. Finally, a new *Request Interpreter* is set and attached to MaaS server to interpret the upcoming run requests to this newly published model. It is argued that provisioning a ready-to-go environment is a complex process in the proposed architecture as it requires extensive collaborations from modelers/researchers and MaaS provider. However, since this process is recorded, the subsequent provisions of the ready-to-go environment can be carried out in minutes.

MOLNs is introduced by Drawert et al. [72] on StochSS infrastructure as a cloud computing appliance to ease the deployment and configuration of computing resources in private, public, or hybrid clouds [75]. The biologists working with StochSS infrastructure can create computing clusters on-demand to execute large-scale computational experiments without having cloud computing skills.

### 3.2.5 Analysis and Visualization

The goal of the *analysis* is to leverage effective techniques (e.g., summarization techniques) for examining the data produced in Data Farming process [73]. *Visualization* aims at optimizing the data exploration process and providing understandable results to decision-makers.

Hüning et al. [43] develop a visual analytics dashboard to provide real-time, visualized data. The real-time visualization of data enables users to optimize simulation models without leaving interface (called Websuite). For example, users are allowed to stop or readjust long-running simulations. In [67], users are also provided with real-time simulation results via WebSocket protocol. In WEEZARD ecosystem [52], users are provided with three distinct views to check the status of simulation jobs: 2D local view, 2D georeferenced view, and 3D view. Furthermore, the “a*synchronous-mode process-request*” procedure, described in Section 3.2.4.2, enables users of WEEZARD to analyze partial results at runtime. The user interface of the SIMaaS infrastructure proposed in [54] provides mechanisms for users to observe the evolution of a simulation runs and its utilized resources. All logs and outputs of a simulation are collected by a logging approach and are visualized at runtime using a *Simulation Info* panel. Ibrahim et al. [57] have presented a framework named SIM-Cumulus for performance analysis of wireless network simulations. Performance is analyzed in terms of execution time and energy footprint. The StochSS infrastructure in [72] leverages a wide range of built-in 3D visualization tools and techniques to represent simulation results. Specifically, the StochSS uses *raster* technique and *ray-tracing* technique to process and animate a model’s 3D information. Whilst the built-in *raster* technique visualizes the surface and internal cross-sections of a spatial model using *domain slicing* feature, the *ray-tracing* technique enables one to explore the concentration fields generated in a spatial model using *volume rendering*. Furthermore, external analytics systems can manipulate simulation results as the StochSS generates the simulation results in a wide range of formats such as StochKit2 format, PyURDME result object, CVS files, and VTK format.

In another attempt, Bozcuoğlu et al. [61] enhance openEASE cloud engine to provide mental simulation for robots. Apart from enabling the execution of computationally expensive simulations, a query interface is developed, which allows users and robots to perform Prolog queries on and reason about simulated experiments.

### 3.2.6 Cross-Analysis of Architectural Solutions

We summarize the architectural solutions identified from the primary studies in Table 17. The *Solution Focus* column in Table 17 has the same classification as categories identified to answer RQ2. The primary studies in each category are focused on different aspects of MSaaS. The studies classified into *Rapid Scenario Prototyping* category focus on different types of interfaces supported by the MSaaS systems. Exposing functionality via web services and web-based interfaces are commonly used strategies. The studies in the *Model Development* category report different ways of developing simulation models. Breaking down complex simulations into smaller services and loose coupling between back-end and front-end services are widely used strategies for model development. The *Design of Experiments* category focuses on how simulation data is stored and how simulation services are provisioned. Dockers and hypervisors are widely used for provisioning simulation services and storing simulation data. The *High-Performance Computing* category discusses high-performance computing requirements in MSaaS and shows how to achieve these requirements. Parallel execution of simulation services, dynamic composition of simulation models, and resource management using middleware are commonly reported strategies for enabling high-performance computing. The studies in *Analysis and Visualization* category report on how to visualize simulation results. Visualization is mainly supported using dashboards and 3D visualization tools.

Table 17. The architectural solutions identified in the primary studies

| **Solution Focus** | **Details** |
|---|---|
| Rapid Scenario Prototyping | Convenient and intuitive interfaces [43] |
| | Browser-based GUI [60] |
| | Google Map HTTP-based APIs [65] |
| | REST front-end interfaces [68] |
| | Exposing functionality via web-services [54] |
| | Command-line interfaces [72] |
| Model Development | Stateless models [1, 50] |
| | Abstracting geoscience model to model configurations [53] |
| | Loosely coupled back-end services [67] |
| | Dividing a complex simulation model into simulated entities [24] |
| | Exposing workflow as a service [48] |
| | Restful web services [42] |
| | Service component simulation using REST web services [1] |
| | Building Model as a Service on IaaS layer [53] |
| Design of Experiments | Signature of resources for input, output, and function [42] |
| | Docket contained images [43] |
| | Databases to store user information and results of simulation experiments [65] |
| | Simulation outputs and related metadata are stored in a centralized database [53] |
| | Simulation information is directly stored in a file system [52] |
| | Storing simulation output on storage based on open standards [68] |
| | Hypervisor-based lightweight virtualization techniques [64] |
| | Docker for virtualizing simulation application [43] |
| | Docker containers for simulation as a service [51, 60] |
| High-Performance Computing | Dynamic resource management for parallelizing simulation jobs [64] |
| | A multi-host network that allows inter-container communications across Docker hosts [9] |
| | Orchestrating Docker containers [60] |
| | Composition of simulation models [47] |
| | A Simulation-as-a-Service framework to communicate with multiple running simulations simultaneously [65] |
| | Asynchronous-mode process-request to enable clients and servers to communication [52] |
| | Model for deploying simulation services in the MSaaS paradigm [14] |
| | Simulation engine for executing simulation models [2] |
| | Parallel running of a model with different configurations and handle concurrent user requests for a model run [53] |
| | Simulating execution requests on different nodes in a processor pool [24] |
| | Execution of multiple services simultaneously for operations over multiple entities [69] |
| | The virtualized infrastructures for VM image and VM entity [46] |
| | Virtualization at runtime [64] |
| | Resources (e.g., Amazon EC2 instances) configuration by customizing StarCluster toolkit with new plug-ins [54] |
| | Docker containers in developing and deploying HLA-based simulation systems [9] |
| | Configurable Windows and Linux images for hosting simulation environments for execution on HCC (High Capacity Computing) platforms [45]<br>Publishing new models into the existing MaaS system [53]<br>Cloud computing resources to ease the deployment and configuration of computing resources in private, public, or hybrid clouds [72, 75] |
| Analysis and Visualization | Visual analytics dashboard [43]<br>Real-time simulation results via WebSocket protocol [67]<br>Analyzing partial results at runtime [54]<br>SIM-Cumulus framework for performance analysis of wireless network simulations [57]<br>3D visualization tools and techniques to represent simulation results [72]<br>OpenEASE cloud engine to provide mental simulation for robots [61] |

## 4. Threats to Validity

This section presents the threats that may have negatively influenced the validity of this study's findings, and the strategies adopted to mitigate them. We follow the guidelines proposed by [76] to report these threats and their corresponding mitigation strategies.

### 4.1 Search Strategy

In SLRs, the chance of missing relevant papers is inevitable. In our review, this threat may have happened during the search strategy and the study selection process. We implemented serval strategies to reduce this issue. We consulted with the existing secondary studies [12, 19], well-known papers, and technical reports produced by NATO [10, 36] to find out as many as possible terms related to MSaaS and their combination. Our search string was finalized after several times pilot-testing different search strings prior to the study selection process. We also checked if excusing the search string on the Scopus search engine could return the well-known papers on MSaaS. One of the limitations of this study is that we only used Scopus to identify relevant papers. As a result, some papers indexed by other databases, such as ACM Digital Library and IEEE Xplore, might have been missed. Our experiences in the previous SLRs [29, 30] show that Scopus well fits with the computer science discipline, as it indexes a vast majority of workshops, conferences, and journals in the computer science discipline. We further took two actions to mitigate this threat. First, we made our search string very broad and comprehensive by including many keywords. Gusenbauer [77] found that among other search engines (e.g., Google Scholar and Web of Science), Scopus is the only search engine that returns more records if a longer search string used. Second, we ensured that simulation- and software architecture-related workshops, conferences, and journals all are indexed by Scopus. Despite this effort, we accept that we might have missed some relevant papers.

### 4.2 Study Selection

Given the first author mainly performed the study selection, there is a possibility of subjective bias in this SLR. We alleviated this bias, to some extent, by taking the following actions: (1) we ensured that the inclusion and exclusion criteria were strictly met in each step of the study selection process. Specifically, the exclusion criterion E3 (See Table 2) was a strict one, as we found a large number of the papers proposing mathematical techniques for simulation models without any discussions about the architecture of underlying infrastructures for the simulation models. (2) During each step of the selection process, we recorded the reason why a study was excluded or included in an Excel spreadsheet file. (3) The last measure that we adopted to reduce the potential threats in the selection process was to hold several internal meetings with the other authors. The spreadsheet file was used as a reference point in the internal meetings to do cross-validation, seek other authors' feedback, and ensure the clarity of the selection process as much as possible.

### 4.3 Data Extraction

Personal bias may occur during the extraction data step (i.e., inconsistent understanding of the selected studies). Before starting the data extraction step, the first author created an Excel spreadsheet to determine what piece of information should be extracted from the selected studies. This form was developed based on the guidelines and standards proposed for SLR [28] and the existing secondary studies [12, 19] and well-known papers. Then, the spreadsheet form was shared with the second author to seek his feedback. After finishing the data extraction, the second and third authors reviewed and validated the extracted data. The spreadsheet file enabled us to identify any misunderstandings, misinterpretations, and confusion in the extracted data and address them collaboratively.

### 4.4 Data Synthesis

Threats in this section stem from the qualitative analysis of the extracted data (i.e., described in Section 2.4). It is because the first author mainly carried out the analysis and classification of the qualitative data. Tómasdóttir et al. [78] assert that this approach increases consistency in coding the qualitative data. However, we accept that this may also be a threat to our review. We took the following mitigation actions to minimize this threat: (1) the first author employed an inclusive coding strategy by following best practices for the qualitative analysis to mitigate this limitation. (2) Two independent persons were invited to review and provide feedback on the early version of the qualitative results. The first one was a researcher who worked on software architectures for simulation systems. The second person was a simulation expert and an active member of the SR2 project. Although we had access to the documents of SR2, the feedback received by the simulation expert helped us avoid any potential misunderstandings of SR2's user requirements. This was critical in reducing the researchers' bias in the interpretations and inferences of the findings obtained in RQ1. (3) Finally, several meetings were organized with the second and third authors to clarify the classifications and interpretations of the qualitative data and exclude any mistakes.

## 5. Discussion and Conclusion

Modelling and Simulation as a Service (MSaaS) paradigm purports to facilitate the development, deployment, and maintenance of M&S applications. We have reviewed the literature on the architectural aspects of underpinning systems supporting MSaaS. This review has identified and analyzed several architectures used for providing MSaaS. We have categorized the identified architectures of underpinning MSaaS systems based on their architectural styles and discussed the potential impacts of each of them in terms of non-functional requirements. We have reported on the strengths and weaknesses of each of the identified architectures. We have also used the requirements of an industrial M&S project (i.e., Strategic Response 2 - SR2) as a criteria-based evaluation framework to assess the state-of-the-art MSaaS architectures. Identifying architectural solutions (e.g., architectural patterns) promoting or supporting MSaaS was another part of this report, in which we have mapped the architectural solutions to the five activities of Data Farming framework developed by NATO. The main findings, lessons learned, and implications of this study can be summarized under four headings:

***Architectural Styles***: Looking at all the reviewed M&S applications, we realize that the layered structure of the cloud heavily influences their architecture. In fact, the layered architecture approach is the dominant style in the reviewed architectures. However, the number of layers and their functionalities are determined based on the main drivers behind the architecture and the underlying infrastructures. We also reviewed the proposed architectures to understand the main quality attributes in the MSaaS paradigm. As shown in Figure 4, interoperability and deployability have the greatest importance in MSaaS applications, followed by cost, performance, scalability, and configurability. Hence, future architectures need to provide mechanisms to satisfy these quality attributes.

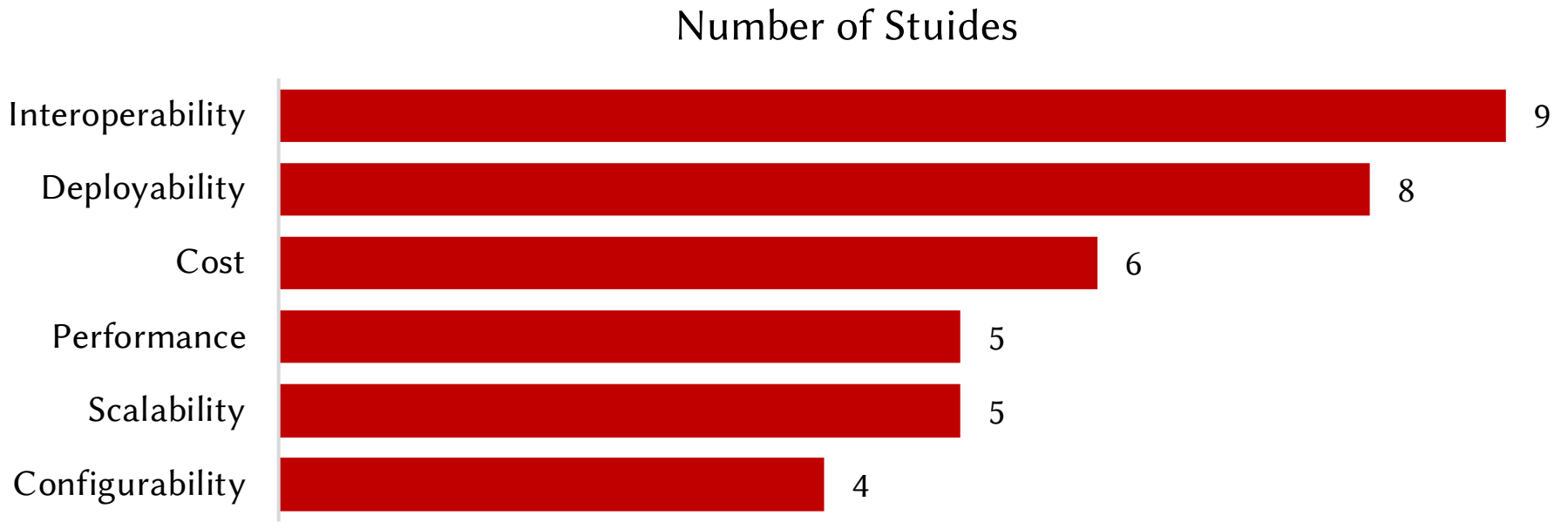

Figure 4. Top six quality attributes achieved or improved by the proposed architectures

***Containerize M&S Applications***: The latest architectures leverage the containerization approach as an alternative to virtualization in order to improve the deployability of M&S applications. It is also argued that whilst a monolithic application can be deployed as a container, containers are ideal for small applications (e.g., microservices-based applications) [79]. According to van den Berg et al. [9], monolithic simulation applications, particularly in the military domain, are dominant, and containerizing this class of systems might pose challenges to their deployment process. We believe when M&S applications are more considering lightweight virtualization (e.g., Docker container) than VMs, there is a need to (re-) design M&S applications in a way that there would be small, loosely coupled simulation applications or services that can be seamlessly set up and deployed. This approach would help to address the various requirements of modern M&S applications, such as parallel execution of simulation models and help to maximize the potential benefits of containers in simulation systems.

***End Users and Interface***: Our study shows that one of the main drivers behind the proposed architectures (See Tables 4, 7, 9, and 11) is to enabling end users to build and analyze simulation models and results in an effective matter. These concerns have a considerable influence on the design of the interface. While an intuitive interface exists on the top of the proposed architectures, our analysis shows that the interface design will get more importance when users with low or without domain knowledge and computational backgrounds working with M&S applications. The (modern) interfaces not only should hide the complexity and execution of simulation runs from end users, but also need to provide mechanisms for building and configuring simulation models as there should be minimum or no effort required by end users. This is partially achievable by giving more flexibility to users to build their own customized models based on specific needs through automating and customizing the simulation models' generation process.

***Architectures for Modern M&S Applications***: In this study, we have reviewed the architectures providing MSaaS from the perspective of SR2's user requirements (i.e., SR2 is considered as a modern M&S application in the military domain). We have presented this information in an easily accessible format (See Tables 6, 9, 12, and 15), so the reader can learn *what* and *how* SR2's user requirements might be addressed by the proposed architectures without having to go through the specification documents for the different architectures. Table 18 groups the reviewed studies based on the SR2's user requirements. It should be noted that the data in Table 18 are generated from the "*Requirements*" columns in Tables 6, 9, 12, and 15. Among the requirements listed in Table 18, **R1**, **R2**, **R6**, and **R7**, are mostly addressed by the reviewed architectures. Table 18 shows that more than half of the reviewed architectures support these requirements. However, four critical SR2's requirements, i.e., **R3**, **R4**, **R5**, and **R8** are rarely addressed as less than 40% of the reviewed studies provide the architectures that satisfy these requirements. The parallel execution of simulations (**R5**) is an attractive feature for the SR2 project. Running each execution independently of others and the containerization approach may help to achieve this goal. However, we still need a new class of approaches and algorithms to schedule simulation jobs for parallel execution and manage data production into a shared resource data farming service [64]. On the other hand, whilst a few studies such as [42, 61] have developed approaches to understand and visualize the significant parts of simulation results after completion of all simulation runs, it is also desirable to monitor, analyze and optimize simulations during their executions (i.e., **R4**). We found only six studies [43, 52, 54, 67-69] that, to some extent, provide such facilities. We believe that this requirement is critical as it allows continuous improvement and evolution of simulations, e.g., one can monitor the progress of a long batch of runs, stop one or more simulation executions, reformulate or modify *design points* based on the feedback gained from the runtime monitoring.

Table 18. Classification of studies by the SR2's user requirements

| Req# | Studies | % | Req# | Studies | % |
|---|---|---|---|---|---|
| **R1** | [42] [47] [48] [49] [50] [2] [51] [53] [9] [59] [64] [1] [68] [70] [71] [72] | **51.6** | **R5** | [45] [52] [53] [59] [65] [64] [1] [24] [56] [57] [69] [72] | **38.7** |
| **R2** | [45] [46] [47] [48] [49] [50] [2] [51] [52] [53] [9] [59] [60] [65] [64] [1] [67] [68] [24] [54] [56] [72] | **70.9** | **R6** | [42] [43] [45] [46] [47] [48] [49] [50] [2] [51] [52] [53] [60] [65] [64] [1] [67] [68] [24] [54] [56] [70] [71] [72] | **77.4** |
| **R3** | [43] [50] [64] [67] [54] | **16.1** | **R7** | [42] [43], [45] [46] [47] [48] [49] [50] [2] [51] [52] [53] [9] [59] [60] [65] [64] [1] [67] [68], [24] [54] [56] [69] [72] | **80.6** |
| **R4** | [43] [52] [67] [68] [54] [69] | **19.3** | **R8** | [52] [53] [59] [64] [24] [56] | **19.3** |

## Acknowledgements

This research has been partially supported by Defence Science and Technology (DST) Group, a part of the Australian Department of Defence (https://www.dst.defence.gov.au/). This research has also been supported by the Cyber Security Research Centre Limited, whose activities are partially funded by the Australian Government's Cooperative Research Centre Program.

## Appendix. Selected Studies

Table 19. An overview of the selected studies.
NF: Not found; NA: Not applicable; C: Conference; W: Workshop; J: Journal

| Ref | Citation Counts | Venue Name | Venue Type | Impact Factor | CORE Ranking | SJR Ranking | H5 Index | H5 Median |
|---|---|---|---|---|---|---|---|---|
| [1] | 9 | Federated Conference on Computer Science and Information Systems | C | NA | NF | NA | 25 | 29 |
| [2] | 8 | Model-driven Approaches for Simulation Engineering | C | NA | NF | NA | NF | NF |
| [9] | 4 | The Journal of Defense Modeling and Simulation | J | No IF | NF | Q2 | 12 | 16 |
| [14] | 1 | The Journal of Defense Modeling and Simulation | J | No IF | NF | Q2 | 12 | 16 |
| [24] | 41 | Simulation Modelling Practice and Theory | J | 2.42 | C | Q1 | 37 | 55 |
| [42] | 7 | Journal of Computational Science | J | 2.50 | NF | Q1 | 36 | 52 |
| [43] | 19 | Agent-Directed Simulation Symposium | C | NA | NF | NA | NF | NF |
| [44] | 2 | Agent-Directed Simulation Symposium | C | NA | NF | NA | NF | NF |
| [45] | 4 | Vehicular Technology Conference | C | NA | B | NA | 44 | 64 |
| [46] | 4 | Symposium on Service Oriented System Engineering | C | NA | NF | NA | 21 | 35 |
| [47] | 2 | Winter Simulation Conference | C | NA | B | NA | 20 | 25 |
| [48] | 15 | Symposium on Service Oriented System Engineering | C | NA | NF | NA | 21 | 35 |
| [49] | 7 | Future Generation Computer Systems | J | 5.76 | A | Q1 | 73 | 110 |
| [50] | 3 | Advances in Engineering Software | J | 4.19 | B | Q1 | 41 | 58 |
| [51] | 4 | INCOSE Italia Conference on Systems Engineering | C | NA | NF | NA | NF | NF |
| [52] | 7 | Environmental Modelling & Software | J | 4.55 | NF | Q1 | 65 | 91 |
| [53] | 33 | Computers, Environment and Urban Systems | J | 3.33 | NF | Q1 | 44 | 72 |
| [54] | 5 | European Conference on Parallel and Distributed Computing | C | NA | A | NA | 21 | 28 |
| [56] | 39 | Annual Simulation Symposium | C | NA | B | NA | NF | NF |
| [57] | 1 | IEEE Access | J | 4.09 | NF | Q1 | 89 | 118 |
| [59] | 7 | International Conference on Industrial Informatics | C | NA | NF | NA | 21 | 33 |
| [60] | 0 | IEEE Biomedical Circuits and Systems Conference | C | NA | NF | NA | 19 | 25 |
| [61] | 11 | IEEE International Conference on Robotics and Automation | C | NA | B | NA | 82 | 113 |
| [64] | 28 | Annals of Telecommunications | J | No IF | NF | Q2 | 20 | 28 |
| [65] | 41 | Simulation Modelling Practice and Theory | J | 2.42 | C | Q1 | 37 | 55 |
| [67] | 1 | Archives of Photogrammetry, Remote Sensing, Spatial Information Sciences | C | NA | NF | NA | 34 | 41 |
| [68] | 0 | International Defense and Homeland Security Simulation Workshop | W | NA | NF | NA | NF | NF |
| [69] | 0 | Symposium on Distributed Simulation and Real-Time Applications | C | NA | NF | NA | 10 | 13 |
| [70] | 4 | Symposium on Model-driven Approaches for Simulation Engineering | C | NA | NF | NA | NF | NF |
| [71] | 1 | Summer Computer Simulation Conference | C | NA | B | NA | NF | NF |
| [72] | 30 | PLoS Computational Biology | J | 4.42 | NF | Q1 | 84 | 115 |